\documentclass[preprint,12pt,nonatbib,nopreprintline]{elsarticle}

\usepackage{graphicx}
\usepackage{amssymb}
\usepackage{lineno}
\usepackage{hyperref}
\usepackage{amsmath}
\usepackage{comment}
\usepackage{multirow}
\usepackage{pdflscape}
\usepackage{adjustbox}
\usepackage{longtable}
\usepackage{geometry}
\usepackage{booktabs}
\usepackage[table]{xcolor} 
\usepackage{changepage}
\usepackage{color,soul}
\usepackage[normalem]{ulem}

\makeatletter
\let\c@author\relax
\makeatother

\usepackage[backend=biber,style=nature,isbn=false,url=false]{biblatex} 
\newcommand{\koff}{k_{\mathrm{off}}}
\newcommand{\kon}{k_{\mathrm{on}}}
\newcommand{\kp}{k_{\mathrm{+}}}
\newcommand{\km}{k_{\mathrm{-}}}

\usepackage{array}
\usepackage{hyperref}
\usepackage{caption}
\usepackage[splitrule]{footmisc} 
\begin{document}
\begin{frontmatter}

\title{A matter of time: Using dynamics and theory to uncover mechanisms of transcriptional bursting}

\author[fref1]{Nicholas C Lammers}
\author[fref1]{Yang Joon Kim \corref{cor2}}
\author[fref2]{Jiaxi Zhao \corref{cor2}}
\author[fref1,fref2,fref3,fref4]{Hernan G Garcia \corref{cor1}}

\fntext[fref1]{Biophysics Graduate Group, University of California at Berkeley, Berkeley, California}
\fntext[fref2]{Department of Physics, University of California at Berkeley, Berkeley, California}
\fntext[fref3]{Department of Molecular and Cell Biology, University of California at Berkeley, Berkeley, California}
\fntext[fref4]{Institute for Quantitative Biosciences-QB3, University of California at Berkeley, Berkeley, California}
\cortext[cor2]{These authors contributed equally to this work.}
\cortext[cor1]{For correspondence: hggarcia@berkeley.edu (HGG)}

\begin{abstract}
Eukaryotic transcription generally occurs in bursts of activity lasting minutes to hours; however, state-of-the-art measurements have revealed that many of the molecular processes that underlie bursting, such as transcription factor binding to DNA, unfold on timescales of seconds. This temporal disconnect lies at the heart of a broader challenge in physical biology of predicting transcriptional outcomes and cellular decision-making from the dynamics of underlying molecular processes. Here, we review how new dynamical information about the processes underlying transcriptional control can be combined with theoretical models that predict 
not only averaged transcriptional dynamics, but also their variability, to formulate testable hypotheses about the molecular mechanisms underlying transcriptional bursting and control.
\end{abstract}

\end{frontmatter}

\begin{refsection}

\section{A disconnect between transcriptional bursting and its underlying molecular processes}

Over the past two decades, new technologies have revealed that transcription is a fundamentally discontinuous process characterized by transient bursts of transcriptional activity interspersed with periods of quiescence. Although electron microscopy provided early hints of bursty transcription \parencite{McKnight1979}, the advent of single-molecule fluorescence \textit{in situ} hybridization (smFISH) \parencite{Femino1998,Raj2008}, was key to establishing its central role in transcription. The single-cell distributions of nascent RNA and cytoplasmic mRNA molecules obtained using this technology provided compelling, if indirect, evidence for the existence and ubiquity of gene expression bursts, and indicated that their dynamics were subject to regulation by transcription factors \parencite{Raj2009,Xu2015b}. These fixed-tissue inferences have been confirmed with new {\it in vivo} RNA fluorescence labeling technologies such as the MS2/MCP \parencite{Bertrand1998} and PP7/PCP systems \parencite{Chao2008}, which directly reveal stochastic bursts of transcriptional activity in living cells in culture and within animals (\autoref{fig:BurstingVsMoleculesTimesScales}A-C) \cite{Golding2005,Chubb2006,Larson2011,Bothma2014}.

What is the role of transcriptional bursting in cellular decision-making? One possibility is that bursty gene expression is intrinsically beneficial, helping (for instance) to coordinate gene expression or to facilitate cell-fate decision-making \parencite{Eldar2010}. Alternatively, bursting may not itself be functional, but might instead be a consequence of key underlying transcriptional processes, such as proofreading  transcription factor identity  \parencite{Grah2020a,Shelansky2020b}. 

Bursting and its regulation are intimately tied to the molecular mechanisms that underlie transcriptional regulation as a whole. In this Review we argue that, to make progress toward predicting transcriptional outcomes from underlying molecular processes, we can start with the narrower question of how the burst dynamics emerge from the kinetics of molecular transactions at the gene locus. To illustrate the importance and challenge of taking kinetics into account, we highlight two interrelated molecular puzzles that arise from new measurements of the dynamics of key transcriptional processes \textit{in vivo}.

First, as illustrated in \autoref{fig:BurstingVsMoleculesTimesScales}D and reviewed in detail in Appendix~\autoref{tab:bursting}, despite qualitatively similar bursty traces from different organisms, bursts unfold across markedly distinct timescales ranging from several minutes~\parencite{Berrocal2018a,Lammers2020}, to tens of minutes~\parencite{Lee2019, Rodriguez2019}, all the way to multiple hours~\parencite{Suter2011}. Is this wide range of bursting timescales across organisms reflective of distinct molecular mechanisms or is it the result of a common set of highly malleable molecular processes? 

Second, recent live imaging experiments have revealed a significant temporal disconnect between transcription factor binding events, which generally last for seconds, and the transcriptional bursts that these events control, which may last from a few minutes to multiple hours. The majority of the molecular processes underlying transcriptional control are highly transient (\autoref{fig:BurstingVsMoleculesTimesScales}E), with timescales ranging from milliseconds to seconds (see Appendix~\autoref{tab:timescale} for a detailed tabulation and discussion of these findings). 

In this Review, we seek to address this second puzzle by surveying key theoretical and experimental advances that, together, should shed light on the molecular origins of transcriptional bursting and transcriptional regulation. We leverage this framework to examine two kinds of molecular-level models that explain how slow burst dynamics could arise from fast molecular processes. Finally, we present concrete experimental strategies based on measuring variability in the timing of bursts that can be used to discern between molecular models of transcriptional bursting. 

Overall, we seek to illustrate how iterative discourse between theory and experiment sharpens our molecular understanding of transcriptional bursting by reformulating cartoon models as concrete mathematical statements. Throughout this Review, we focus on illustrative recent experimental and theoretical efforts; we therefore do not attempt to provide a comprehensive review of the current literature (see \parencite{Coulon2013,Lenstra2016,Wang2019a,Rodriguez2020,Wong2020,Phillips2019} for excellent reviews).

\begin{figure}
\centering
\makebox[\textwidth][c]{\includegraphics[width=190mm]{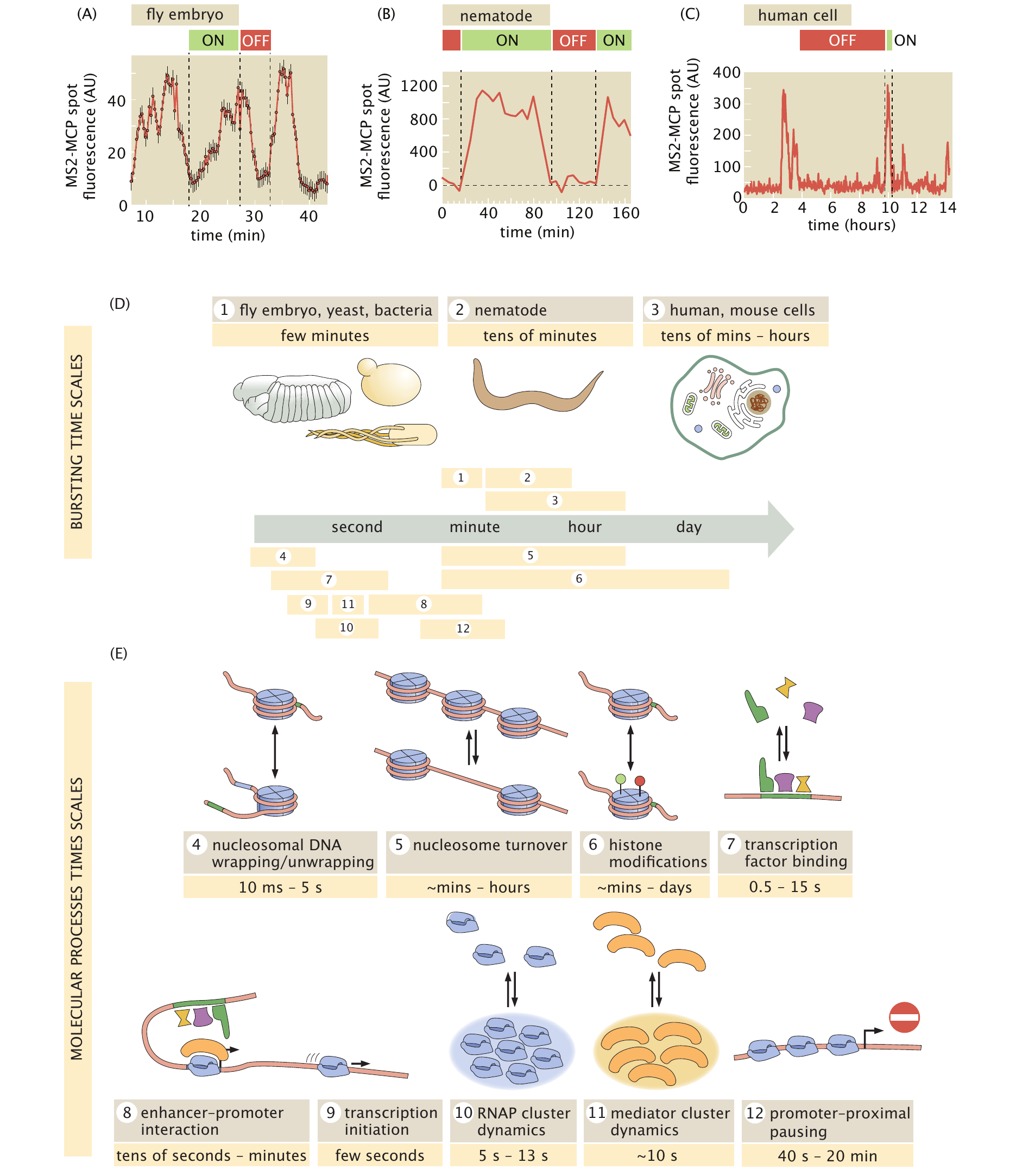}}
\caption{{\bf Separation of timesscales between transcriptional bursting and its underlying molecular processes.} See caption in next page.}
\label{fig:BurstingVsMoleculesTimesScales}
\end{figure}
\setcounter{figure}{0}

\begin{figure}[t]
\caption{{\bf Separation of timescales between transcriptional bursting and its underlying molecular processes.} 
{\bf (A,B,C)} Transcriptional bursting in (A) an embryo of the fruit fly {\it Drosophila melanogaster}, (B) the nematode {\it Caenorhabditis elegans}, and (C) human cells. {\bf (D)} In these and other organisms, bursting dynamics (average period of ON and OFF) span a wide range of timescales from a few minutes to tens of hours. {\bf (E)} Timescales of the molecular processes behind transcription range from fast seconds-long transcription factor binding to slower histone modifications, which may unfold across multiple hours or days. A detailed summary of measurements leading to these numbers, including references, is provided in Appendix~\autoref{tab:bursting} and Appendix~\autoref{tab:timescale}. (A, adapted from \parencite{Lammers2020}; B, adapted from \parencite{Lee2019}; C, adapted from \parencite{Rodriguez2019}).}
\label{fig:BurstingVsMoleculesTimesScales}
\end{figure}

\section{The two-state model: a simple quantitative framework for bursting dynamics}

\begin{figure}[hbt!]
\includegraphics[width=140mm]{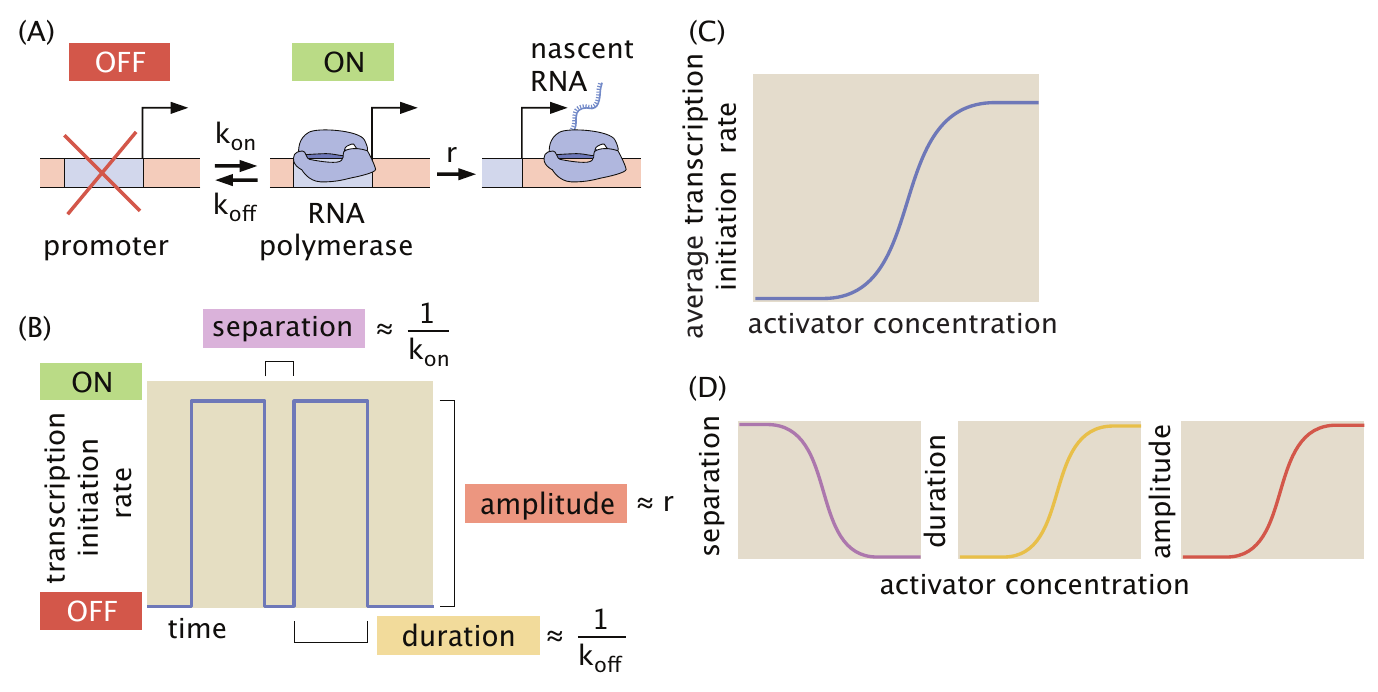}
\centering
\caption{{\bf The two-state model of transcriptional bursting}. {\bf (A)} A two-state model of transcriptional bursting by a promoter switching between ON and OFF states. {\bf (B)} Mapping the bursting parameters $\kon$, $\koff$, and $r$ to burst duration, separation, and amplitude, respectively. {\bf (C)} The action of an activator results in an increase in the average rate of transcription initiation. {\bf (D)} In the two-state model, this upregulation can be realized by decreasing burst separation, increasing burst duration, increasing burst amplitude, or any combination thereof.}
\label{fig:Bursting_features}
\end{figure}

To elucidate the disconnect between molecular timescales and transcriptional bursting, we invoke a simple and widely used model of bursting dynamics: the two-state model of promoter switching. While the molecular reality of bursting is likely more complex than the two-state model suggests \parencite{Corrigan2016,Zoller2015,Morrison2020}, there is value in examining where this simple model breaks down. This model posits that the promoter can exist in two states: a transcriptionally active ON state and a quiescent OFF state (Figure~\ref{fig:Bursting_features}A). The promoter stochastically switches between these states with rates $\kon$ and $\koff$, and loads new RNA polymerase II (RNAP) molecules at a rate $r$ when in the ON state \parencite{Wang2019a,Sanchez2013,Boeger2015,Munsky2015}. Figure~\ref{fig:Bursting_features}B shows a hypothetical activity trace for a gene undergoing bursty expression, where a burst corresponds to a period of time during which the promoter is in the ON state. The average burst duration, amplitude and separation are given by $1/k_{off}$, $r$ and $1/k_{on}$, respectively.

Because the {\it instantaneous} transcription initiation rate during a burst is $r$ and zero otherwise, the \textit{average} initiation rate is equal to $r$ times the fraction of time the promoter spends in this ON state $p_{on}$,
\begin{equation} \label{eq:mRNA_rate}
    \Big \langle \mbox{initiation rate} \Big \rangle  = r \, p_{on},
\end{equation}
where brackets indicate time-averaging. As shown in \ref{app:two_state_model}, in steady state, $p_{on}$ can be expressed as a function of the transition rates $\kon$ and $\koff$:
\begin{equation}\label{eq:pon_sol}
    p_{on} = {k_{on} \over k_{on} + k_{off}}.
\end{equation}
Plugging this solution into Equation~\ref{eq:mRNA_rate} results in the average rate of mRNA production as a function of the bursting parameters given by
\begin{align}\label{eq:mean_rate_2state}
    \Big \langle \mbox{initiation rate} \Big \rangle &=  \underbrace{r}_{\substack{\mbox{transcription rate} \\ \mbox{in ON state}}} \underbrace{\frac{\kon}{\kon + \koff}}_{\substack{\mbox{probability of} \\ \mbox{ON state}}}.
\end{align}

\autoref{eq:mean_rate_2state} shows that, within the two-state model, transcription factors can influence the mean transcription rate by modulating any one of the three burst parameters (or a combination thereof). For example, consider an activator that can increase the mean transcription rate (\autoref{fig:Bursting_features}C) by decreasing $\koff$, increasing $\kon$ or $r$, or any combination thereof (\autoref{fig:Bursting_features}D). Both live-imaging measurements and smFISH have revealed that the vast majority of transcription factors predominantly modulate burst separation by tuning $\kon$ \parencite{Fukaya2016c,Lammers2020,Berrocal2018a,Bothma2014,Zoller2018,Desponds2016e,Xu2015b,Lionnet2011}. There are also examples of the control of burst amplitude and duration, however \parencite{Zoller2018,Falo-Sanjuan2019,Lee2019}. 

Yet although experiments have identified {\it which} bursting parameters are under regulatory control, the question of {\it how} this regulation is realized at the molecular level remains largely open (with a handful of notable exceptions in bacteria \parencite{Chong2014a}, yeast \parencite{Donovan2019}, and mammalian cell culture \parencite{Nicolas2017}). This is because the two-state model is a \textit{phenomenological} model: we can use it to quantify burst dynamics without making any statements about the molecular identity of the burst parameters. Nonetheless, by putting hard numbers to bursting and identifying which parameter(s) are subject to regulation, this framework constitutes a useful quantitative tool to formulate and test hypotheses about the molecular mechanisms underlying transcriptional control.

For instance, consider the observation that many activators modulate burst separation. A simple way to explain this fact is to posit that transitions between the ON and OFF states reflect the binding and unbinding of individual factors to regulatory DNA. Here, $k_{off}$ would be the activator DNA-unbinding rate and $k_{on}$ would be a function of activator concentration $[A]$,
\begin{equation} \label{eq:activator_binding_rate}
    k_{on}([A]) = [A] k^b_0,
\end{equation}
where $k^b_0$ is the rate constant for activator binding.

A recent study in yeast lent credence to this picture, finding that activator affinity ($k^u$) might indeed play a role in dictating burst duration \parencite{Donovan2019}. However, for most genes and organisms surveyed so far, the two-state model indicates that transcription factor unbinding alone cannot set the timescale for bursting: if $\koff$ were an activator unbinding rate, then it would be on the order of $1~\text{s}^{-1}$ (\autoref{fig:BurstingVsMoleculesTimesScales}D~and~E, box 7). Yet, measurements of burst duration reveal that $\koff$ must be orders of magnitude smaller ($\lesssim 0.01 \text{ s}^{-1}$, \autoref{fig:BurstingVsMoleculesTimesScales}D). Thus, the two-state model lends a quantitative edge to the disconnect in \autoref{fig:BurstingVsMoleculesTimesScales}, reaffirming that transcriptional bursting is unlikely to be solely determined by the binding kinetics of the transcription factors that regulate it. We must therefore extend our simple two-state framework to incorporate molecular mechanisms that allow rapid transcription factor binding and transcriptional bursts that are orders of magnitude slower.

\section{Bridging the timescale gap: kinetic traps and rate-limiting steps} \label{sec:bridge_gap}

Recent works have considered kinetic models of transcription that describe transition dynamics between distinct microscopic transcription factor binding configurations. These models make it possible to investigate how molecular interactions facilitate important behaviors such as combinatorial regulatory logic, sensitivity to changes in transcription factor concentrations, the specificity of interactions between transcription factors and their targets, and transcriptional noise reduction \parencite{Shelansky2020b,Grah2020a,Id2018,Li2018b,Scholes2017,Estrada2016a,Desponds2020a}.

We illustrate how these kinetic models can shed light on the disconnect between the timescales of transcription factor binding and bursting using the activation of the {\it hunchback} minimal enhancer by Bicoid in the early fruit fly embryo as a case study \parencite{Gregor2007a,Estrada2016a,Park2019,Id2018,Eck2020,Desponds2016e}. Recent {\it in vivo} single-molecule studies have revealed that Bicoid specifically binds DNA in a highly transient fashion ($\sim1-2$~s) \parencite{Mir2017,Mir2018b}, suggesting that Bicoid binding cannot dictate the initiation and termination of \textit{hunchback} transcriptional bursts, which happen over minutes \parencite{Desponds2016e}. We seek molecular models that recapitulate two key aspects of bursting: (1) the emergence of effective ON and OFF transcriptional states, and (2) ``slow'' ($>$1~min) fluctuations between these states. We sketch out the mathematical basis of these efforts and review key results below; more detailed calculations can be found in \ref{app:molecular_model_calc}.

Following \parencite{Id2018}, we consider a simple activation model featuring an enhancer with identical activator binding sites. While the full model for the \textit{hunchback} minimal enhancer consists of six binding sites, we first use a simpler version with three binding sites to introduce key features of our binding model before transitioning to the more realistic six binding sites version when discussing our results. We capture the dynamics of activator binding and unbinding at the enhancer by accounting for the transitions between all possible binding configurations (\autoref{fig:2models}A). Our assumption of identical activator binding sites leads to two simplifications: (1) the same rate, $k_{i,j}$ governs the switching from any configuration with $i$ activators bound to any configuration with $j$ bound and, (2) all binding configurations with the same number $n$ of activators bound have the same rate of transcription, $r_n = r_0 \, n$, which we posit to be proportional to the number of bound activators. As a result we need not track specific binding configurations and may condense the full molecular representation in \autoref{fig:2models}A into a simpler four-state chain-like model with one state for each possible value of $n$ (\autoref{fig:2models}B). 

\begin{figure}
\centering
\makebox[\textwidth][c]{\includegraphics[width=190mm]{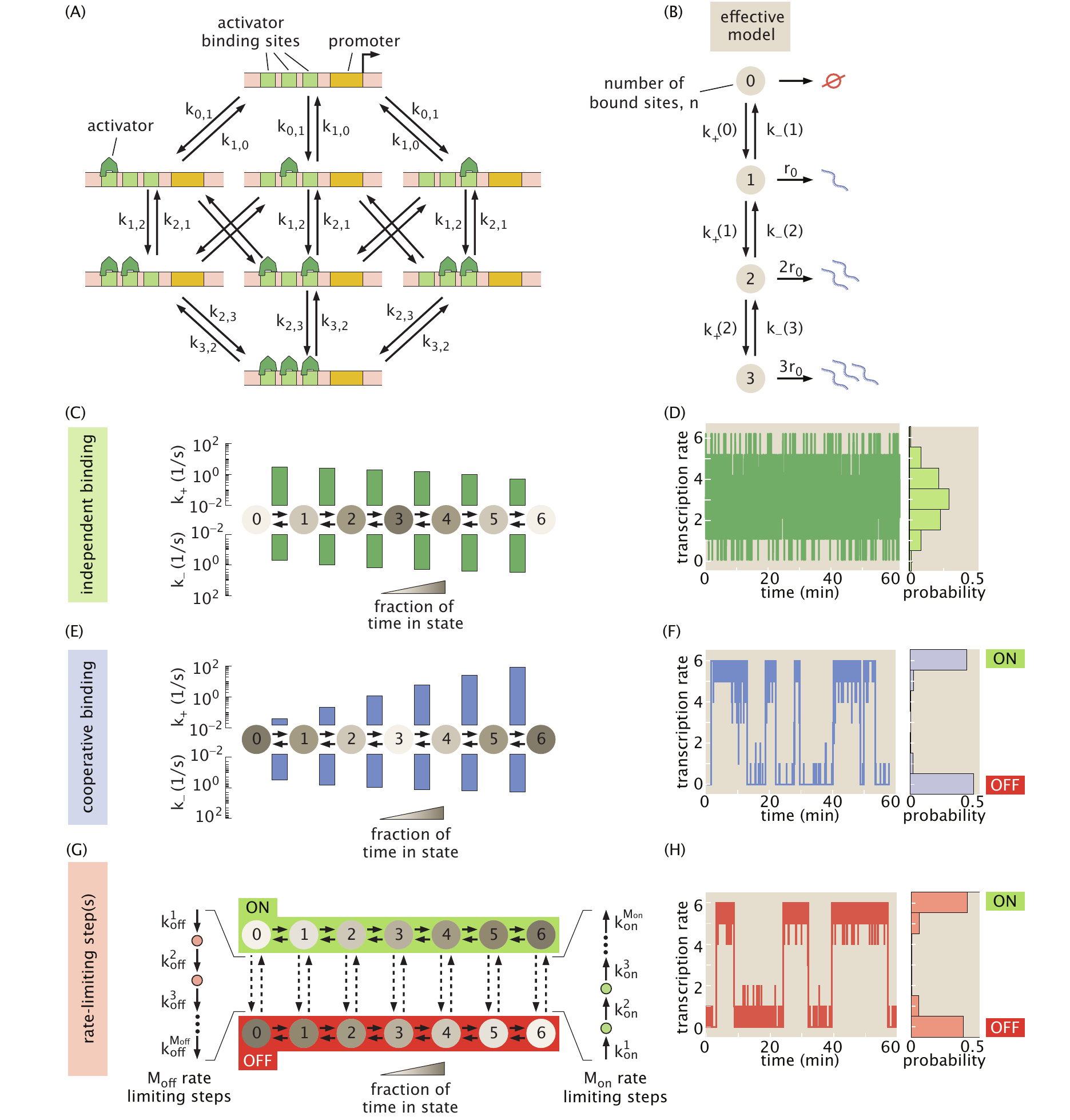}}
\caption{\textbf{Using theoretical models to understand the origin of ON/OFF bursting dynamic}. See caption in the next page.}
\label{fig:2models}
\end{figure}
\setcounter{figure}{2}

\begin{figure}[t]
\caption{\textbf{Using theoretical models to understand the origin of ON/OFF bursting dynamic}.
\textbf{(A)} Model with three activator binding sites. The transition rates between states with $i$ and $j$ activators are given by $k_{i,j}$. \textbf{(B)} The model in (A) can be simplified to an effective four-state chain model in which each state corresponds to a certain number of bound molecules and the transcription rate is proportional to the number of bound activators.
\textbf{(C)} Independent activator binding model with effective binding and unbinding rates plotted above and below, respectively. Shading indicates the fraction of time that the system spends in each state. \textbf{(D)} Stochastic simulations indicate that rapid activator binding alone drives fast fluctuations about a single transcription rate. \textbf{(E)} Cooperative binding model in which already-bound activators enhance the binding rate of further molecules. \textbf{(F)} Simulation reveals that cooperativity can cause the system to exhibit bimodal rates of transcription and slow fluctuations between effective ON and OFF states. \textbf{(G)} Rate-limiting step model in which several molecular steps can connect a regime where binding is favored (ON) and a realization where binding is disfavored (OFF). \textbf{(H)} Simulations demonstrate that rate-limiting steps can lead to bimodal transcriptional activity reminiscent of transcriptional bursting. Simulation results were down-sampled to a resolution of 0.5 s to ensure plot clarity in D, F, and H. Scripts used to generate plots in (D), (F), and (H) are available on GitHub \parencite{Lammers2020c}.
(Parameters: C, D, $k^b=k^u=0.5 \text{s}^{-1}$; E, F, $k^b= 0.004 \text{ s}^{-1}$, $k^u=0.5 \text{ s}^{-1}$; and $\omega=6.7$;
G,H, $k^u_{on}=k^u_{off}= 0.5~\text{ s}^{-1}$, $k^b_{off}= 0.01~\text{ s}^{-1}$, $k^b_{on}= 21~\text{ s}^{-1}$, $M_{off}=1$, $M_{on}=2$, $k_{off}^1=0.0023$, $k_{on}^1=k_{on}^2=0.0046~\mbox{ s}^{-1}$.)}
\label{fig:2models}
\end{figure}

Transitions up and down the chain in \autoref{fig:2models}B are governed by the effective binding and unbinding rates $\kp(n)$ and $\km(n)$. To calculate these rates from the microscopic transition rates $k_{i,j}$, consider, for example, that there are three possible ways of transitioning from the 0 state to the 1 state, each with rate $k_{0,1}$. Thus, the effective transition rate between states 0 and 1 is given by $3 k_{0,1}$. More generally, in the effective model, activator binding rates are
\begin{equation}
    \kp(n) = (N-n) k_{n,n+1}  \label{eq:ind_on_rates_main},
\end{equation}
where $n$ indicates the current number of bound activators and $N$ is the total number of binding sites. Similarly, activator unbinding rates are given by
\begin{equation}
    \km(n) = n k_{n,n-1} \label{eq:ind_off_rates_main}.
\end{equation}
These transition rates allow us to generalize to the more realistic enhancer with six binding sites. We first examine a system in which activator molecules bind and unbind independently from each other (\autoref{fig:2models}C). There are only two unique microscopic rates in this system: activator molecules bind at a rate $k_{i,i+1} = k^{b} = k^{b}_0 [A]$, with $[A]$ being the activator concentration and $k^{b}_0$ the binding rate constant, and unbind at a rate $k_{i,i-1} = k^u$. We fix the unbinding rate $k^u=0.5~s^{-1}$ to ensure consistency with recent experimental measurements of Bicoid in \parencite{Mir2017,Mir2018b}. For simplicity, we also set $k^b=0.5~\mbox{s}^{-1}$ (see \ref{app:diffusion_kb} for details). 

To gain insight into the model's transcriptional dynamics, we employ stochastic simulations based on the Gillespie Algorithm \parencite{Gillespie1977}; however a variety of alternative analytic and numerical approaches exist \parencite{Estrada2016a,Eck2020,Scholes2017}. Our simulations reveal that independent binding leads to a unimodal output behavior in which the transcription rate fluctuates rapidly about a single average (\autoref{fig:2models}D). This result is robust to our choices of $k^b$ or $k^u$, as well as the number of binding sites in the enhancer (\ref{app:independen_binding_no_bursting}). The observed lack of slow, bimodal fluctuations leads us to conclude that the independent binding model fails to recapitulate transcriptional bursting.

One way to extend the independent binding model is to allow for cooperative protein-protein interactions between activator molecules \parencite{Uzman2002}. Specifically, we consider a model where bound activator molecules act to catalyze the binding of additional activators. Here, the activator binding rate is increased by a factor $\omega$ for every activator already bound, leading to
\begin{equation}
    k_{i,i+1} = k^b \omega^i. \label{eq:coop_binding_micro}
\end{equation}
Because we assume that activator unbinding still occurs independently, the effective unbinding rates remain unchanged (\autoref{eq:ind_off_rates_main}). 

Stochastic simulations of the cooperative binding model in \autoref{fig:2models}F reveal that the output transcription rate now takes on an all-or-nothing character, fluctuating between high and low values that act as effective ON and OFF states. Further, our simulation indicates that these emergent fluctuations are quite slow (0.13~transitions/min for the system
shown), despite fast activator binding kinetics. Both of these phenomena result from large imbalances between $\kp(n)$ and $\km(n)$ that act as ``kinetic traps''.

Consider the case with five bound activators. If $\kp(5) \gg \km(5)$, then the enhancer is much more likely to bind one more activator molecule and move to state six than to lose an activator and drop to state four. For instance, if $\kp(5)/\km(5)=23$ (\autoref{fig:2models}F), then the system will on average oscillate back and forth between states five and six 23 times before it finally passes to state four. While it is possible to generate this kind of trap without cooperativity at one end of the chain or the other by tuning $k^b$, cooperative interactions are needed to simultaneously achieve traps at both ends.

While we focused on binding-mediated cooperativity here, we note that all results presented above hold equally well for the unbinding-mediated case where cooperative interactions between bound molecules stabilize binding by reducing $k_{i,i-1}$ while maintaining $k_{i,i+1}$ unchanged. As discussed in \ref{app:koff_cooperativity}, our analysis of this unbinding-mediated cooperativity scenario makes the intriguing additional prediction that rapid ($\sim 1~\text{s}$) activator dwell times inferred from \textit{in vivo} experiments could mask the existence of rare long-lived ($\geq 10~\text{s}$) binding events that, despite their infrequency, play a key role in driving slow transcriptional burst dynamics. Finally, it is important to note that the phenomenon of emergent slow fluctuations is not limited to activator binding: cooperative interactions in fast molecular reactions elsewhere in the transcriptional cycle, such as in the dynamics of pre-initiation complex assembly, could, in principle, also induce slow fluctuations. 

Inspired by the MWC model of protein allostery \parencite{Eck2020,Marzen2013}, a second way to bridge the timescale gap between activator binding and transcriptional bursting is to posit two distinct system configurations: an ON configuration where binding is favored ($k^b\gg k^u$) and an OFF configuration that is less conducive to binding ($k^b\ll k^u$). From any of the seven binding states, this system can transition from OFF to ON by traversing $M_{on}$ slow steps, each with rate $k_{on}^i\ll k^u$, where $i$ is the step number (\autoref{fig:2models}G). Similarly, transitions from ON to OFF are mediated by $M_{off}$ steps with rates given by $k_{off}^i$. Stochastic simulations indicate that this system yields bimodal transcription that fluctuates between high and low activity regimes on timescales set by the rate-limiting molecular steps (\autoref{fig:2models}H). Thus, as long as these steps induce a sufficiently large shift in activator binding ($k^b$), the rate-limiting step model reconciles rapid activator binding with transcriptional bursting. 

\autoref{fig:BurstingVsMoleculesTimesScales}B suggests candidates for these slow molecular steps. For example, the ON state in \autoref{fig:2models}G could correspond to an open chromatin state that favors binding while the OFF state could indicate that a nucleosome attenuates binding such that $M_{on}=M_{off}=1$. Our model also allows multiple distinct rate-limiting steps. For instance, chromatin opening could require multiple histone modifications ($M_{on}\geq2$, $M_{off}=1$), or chromatin opening may need to be followed by enhancer-promoter co-localization to achieve a high rate of transcription ($M_{on}=2$, $M_{off}=1$).

Although they are not the only possible models, the cooperativity and rate-limiting step scenarios discussed above represent two distinct frameworks for thinking about how slow processes like bursting can coincide with, and even arise from, rapid processes like activator binding. The next challenge in identifying the molecular processes that drive transcriptional bursting is to establish whether these models make experimentally distinguishable predictions.

\section{Using bursting dynamics to probe different models of transcription}
\label{sec:pt_times}

\begin{figure}
\centering
\includegraphics[width=140mm]{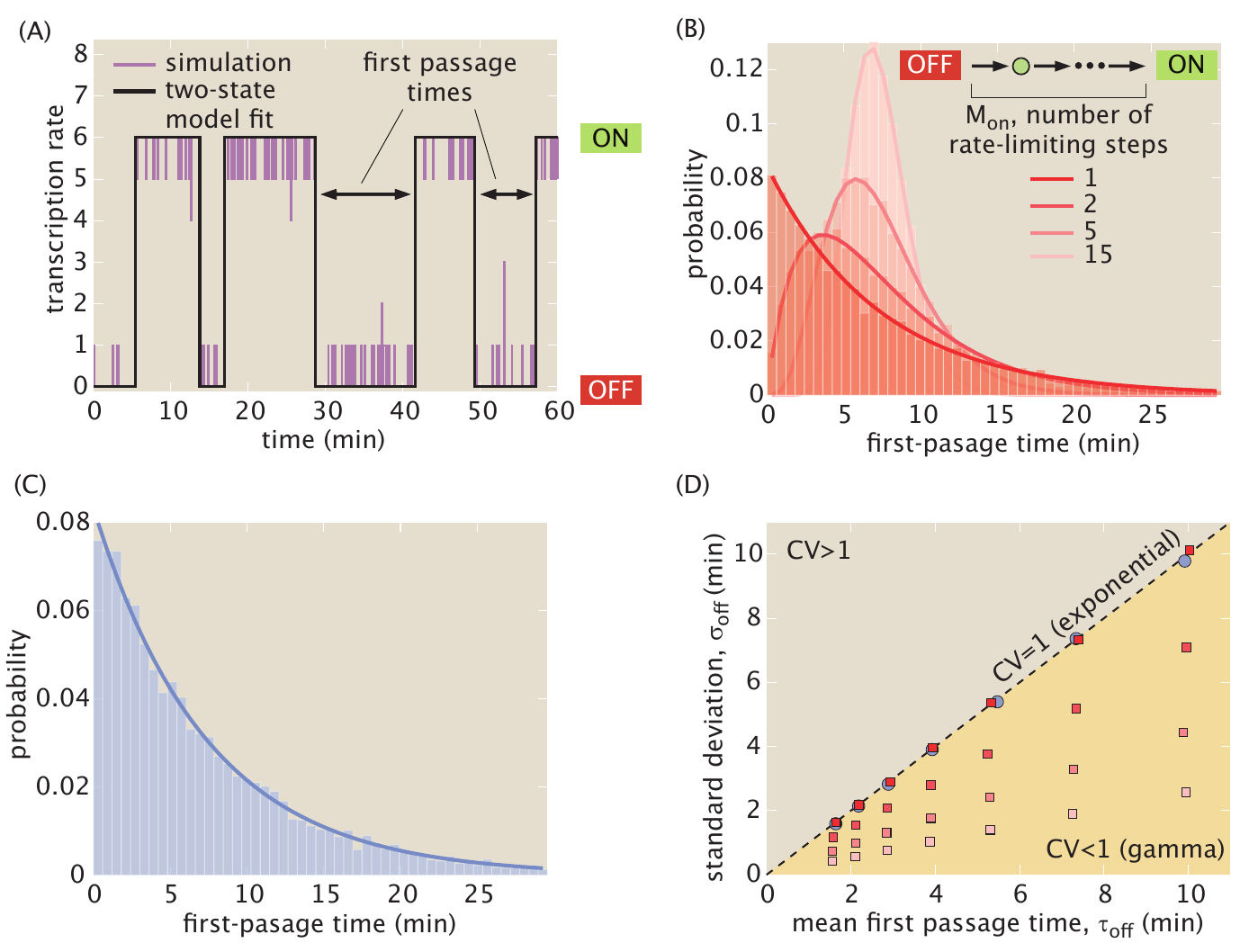}
\caption{\textbf{Using first-passage time distributions to discriminate between models of transcriptional bursting}. \textbf{(A)} The outcome of stochastic simulations like those in \autoref{fig:2models}D, F, and G (purple) is fit to a two-state model (black) and the first-passage times out of the OFF state are measured. \textbf{(B)} First-passage times for the rate-limiting step model as a function of the number of rate-limiting steps $M_{on}$ calculated using stochastic simulations. A single step results in an exponential distribution, but distributions break from this behavior when more steps are added, yielding increasingly peaked gamma distributions. \textbf{(C)} In contrast, first-passage times for the cooperative binding model follow an exponential distribution. \textbf{(D)} Standard deviation as a function of mean first-passage time for various parameters choices of the cooperative binding (blue) and the rate-limiting step models (red, with color shading indicating the $M_{on}$ values considered in (B)). Distributions with $CV=1$, such as the exponential distribution, fall on the line of slope one while gamma distributions, with $CV<1$, fall in the region below this line.  Scripts used to generate plots in (A), (B), (C), and (D) are available on GitHub \parencite{Lammers2020c}. (Parameters: B, $k^u_{on}=k^u_{off}= 0.5~\text{s}^{-1}$, $k^b_{off}= 0.01~\text{s}^{-1}$, $k^b_{on}= 21~\text{s}^{-1}$, $M_{off}=1$, $k_{off}^1=0.0023~\mbox{s}^{-1}$, $k_{on}^i=M_{on} 0.0023~\mbox{s}^{-1}$; C, $k^b= 0.004 \text{s}^{-1}$, $k^u= $$\textcolor{green}{=0.5~\text{s}^{-1}}$, and $\omega=6.7$;
D (rate-limiting-step model), $k^u_{on}=k^u_{off}= 0.5~\text{s}^{-1}$, $k^b_{off}= 0.01~\text{s}^{-1}$, $k^b_{on}= 21~\text{s}^{-1}$, $M_{off}=1$, $\tau_{off}\in[1.6,10.9]~\text{min}$, $k_{off}^1=1/\tau_{off}~\mbox{s}^{-1}$, $k_{on}^i=M_{on} / \tau_{off}~\mbox{s}^{-1}$ ; D (cooperativity model), $k^b \in [0.01,0.003]~\text{ s}^{-1}$, $k^u=0.5 \text{ s}^{-1}$; and $\omega \in[4.5,7.4]$).}
\label{fig:first_passage_times}
\end{figure}

While we cannot yet directly observe the microscopic reactions responsible for bursting in real time, these processes leave signatures in transcriptional dynamics that may distinguish molecular realizations of bursting such as those of our cooperative binding (\autoref{fig:2models}E) and rate-limiting step (\autoref{fig:2models}G) models.
Inspired by \parencite{Yildiz2004,Suter2011c,Zoller2015,Dufourt2018a}, we examine whether the distribution of observed burst separation times (\autoref{fig:first_passage_times}A) distinguishes between these two models. In keeping with literature convention, we refer to these separation times as \textit{first-passage times} from OFF to ON.

The variability in reactivation times provides clues into the number of hidden steps in a molecular pathway. For instance, suppose that bursts are separated by an average time $\tau_{off}=1/\kon$, as defined in the two-state model in \autoref{fig:Bursting_features}A~and~B.  If there is only a single rate-limiting molecular step in the reactivation pathway ($M_{on}=1$ in \autoref{fig:2models}G), then the first-passage times will follow an exponential distribution (\autoref{fig:first_passage_times}B) such that the variability, defined as the standard deviation ($\sigma_{off}$), will simply be equal to the mean ($\tau_{off}$). Now, consider the case where two distinct molecular steps, each taking an average $\tau_{off}/2$, connect the OFF and ON states ($M_{on}=2$). To calculate the variability in the time to complete \textit{both} steps and reactivate, we need to add the variability of each step in quadrature:
\begin{equation}
    \sigma_{off} = \sqrt{\Big(\frac{\tau_{off}}{2}\Big)^2 + \Big(\frac{\tau_{off}}{2}\Big)^2} = \frac{\tau_{off}}{\sqrt{2}}.
\end{equation}
More generally, in the simple case in which each step has the same rate, given an average first-passage time of $\tau_{off}$, the variability in the distribution of measured first-passage times will decrease as the number of rate-limiting steps, $M_{on}$, increases following
\begin{equation}
    \sigma_{off}(M_{on}) = \frac{\tau_{off}}{\sqrt{M_{on}}} \label{eq:sigma_vs_steps}. 
\end{equation}

As predicted by \autoref{eq:sigma_vs_steps}, increasing the rate-limiting step number reduces the width of the distribution for the rate-limiting step model obtained from stochastic simulations, shifting passage times from an exponential distribution when $M_{on}=1$ to increasingly peaked gamma distributions when $M_{on}>1$ (\autoref{fig:first_passage_times}B). 

Based on these results, since the fluctuations between high- and low-activity regimes reflect transitions through many individual binding states in the cooperative binding model (\autoref{fig:2models}E), we might also expect this model to exhibit non-exponential first passage times.  Instead, the first-passage times are exponentially distributed (\autoref{fig:first_passage_times}C). This result is consistent with earlier theoretical work that examined a chain model similar to ours and found that sufficiently large reverse rates ($k^u$ in our case) cause first-passage time distributions to exhibit approximately exponential behavior \parencite{Bel2010}. 

The coefficient of variation ($CV=\sigma_{off}/\tau_{off}$)
provides a succinct way to summarize the shape of passage time distributions for a wide range of model realizations. \autoref{fig:first_passage_times}D plots $\sigma_{off}$ against $\tau_{off}$ for each of the model architectures considered in \autoref{fig:first_passage_times}B~and~C for a range of different $\tau_{off}$ values. Points representing distributions with $CV=1$ will fall on the line with slope one and points for distributions with $CV<1$ will fall below it. We see that both the cooperative binding model and the single rate-limiting step model have $CV$ values of approximately one for a wide range of $\tau_{off}$ values, consistent with exponential behavior. Conversely, all models with multiple rate-limiting steps have slopes that are significantly less than one. 

Thus, by moving beyond experimentally measuring average first-passage time for a given gene and examining its {\it distribution}, it is possible to rule out certain molecular mechanisms. For example, a non-exponential distribution would be evidence against the cooperative binding and single rate-limiting step models (see \ref{app:stoch_sim} and \ref{app:fp_calc} for details about stochastic simulations and first-passage time calculations).
While these conclusions are specific to the models considered here, the general approach of invoking the distributions rather than means and using stochastic simulations to derive expectations for different models can be employed to discriminate between molecular hypotheses in a wide variety of contexts. Indeed, the examination of distributions has been revolutionary throughout biology by making it possible to, for example, reveal the nature of mutations \parencite{Luria1943}, uncover mechanisms of control of transcriptional initiation \parencite{Sanchez2011} and elongation \parencite{Serov2017,Ali2020}, measure translational dynamics \parencite{Cai2006}, and even count molecules \parencite{Rosenfeld2005}.

Note that, while appropriate for {\it qualitatively} estimating the order of magnitude of  bursting timescales, raw fluorescence measurements from MS2 and PP7 experiments such as those in \autoref{fig:BurstingVsMoleculesTimesScales}A-C do not directly report on the promoter state. Rather, the signal from these experiments is a convolution of the promoter state and the dwell time of each nascent RNA molecule on the gene body \parencite{Lammers2020}. As a result, inference techniques like those developed in \parencite{Lammers2020,Corrigan2016} are often required to infer underlying burst parameters and promoter states that can be used to estimate first-passage time distributions. Other techniques, such as measuring the short-lived luminescent signal from reporters \parencite{Zoller2015}, have also successfully estimated first-passage times.

The first-passage time analyses discussed here are just one of an expansive set of approaches to determining the best model to describe experimental data. For instance, direct fits of models to experimental time traces could be used to identify the most appropriate model (see, e.g. \parencite{Corrigan2016,Silk2014}). A discussion of this and other approaches falls beyond the scope of this work, but we direct the reader to several excellent introductions to elements of this field \parencite{Silk2014,Sivia2006,Mehta2019,Devilbiss2017}. 

\section{Conclusions}

The rapid development of live-imaging technologies has opened unprecedented windows into \textit{in vivo} transcriptional dynamics and the kinetics of the underlying molecular processes. We increasingly see that transcription is complex, emergent, and---above all---highly dynamic, but experiments alone still fail to reveal how individual molecular players come together to realize processes that span a wide range of temporal scales, such as transcriptional bursting.

Here we have argued that theoretical models can help bridge this crucial disconnect between single-molecule dynamics and emergent transcriptional dynamics. By committing to mathematical formulations rather than qualitative cartoon models, theoretical models make concrete quantitative predictions that can be used to generate and test hypotheses about the molecular underpinnings of transcriptional control. We have also shown how, although different models of biological phenomena might be indistinguishable in their averaged behavior, these same models often make discernible predictions at the level of the distribution of such behaviors. 

Moving forward, it will be critical to continue developing models that are explicit about the kinetics of their constituent molecular pieces, as well as statistical methods for connecting these models to \textit{in vivo} measurements in an iterative dialogue between theory and experiment. In particular, robust model selection frameworks are needed to navigate the enormous space of possible molecular models for transcriptional control. Such theoretical advancements will be key if we are to synthesize the remarkable experimental findings from recent years into a truly  mechanistic understanding of how transcriptional control emerges from the joint action of its molecular components.

\section{Acknowledgements}

We are grateful to Simon Alamos, Lacramioara Bintu, Xavier Darzacq, Jonathan Desponds, Hinrich Boeger, Michael Eisen, Julia Falo-Sanjuan, Anders Hansen, Jane Kondev, Daniel Larson, Tineke Lenstra, Jonathan Liu, Mustafa Mir, Felix Naef, Rob Phillips, Alvaro Sanchez, Brandon Schlomann, Mike Stadler, Meghan Turner and Aleksandra Walczak for useful discussions and comments on the manuscript. However, any errors and omissions are our own. HGG was supported by the Burroughs Wellcome Fund Career Award at the Scientific Interface, the Sloan Research Foundation, the Human Frontiers Science Program, the Searle Scholars Program, the Shurl \& Kay Curci Foundation, the Hellman Foundation, the NIH Director's New Innovator Award (DP2 OD024541-01), and an NSF CAREER Award (1652236).

\newpage
\appendix

\newgeometry{textwidth=16cm}
\section{Literature summary of timescales of transcriptional bursting and associated molecular processes}

In this section, we present a survey of timescales observed for transcriptional bursting across a broad swath of organisms (Appendix~\autoref{tab:bursting}). Further, we review {\it in vivo} and {\it in vitro} measurements that have revealed the timescales of the molecular transactions underlying transcription and its control.

Recent technological advances such as single-molecule tracking, live-cell imaging, and a variety of high-throughput sequencing methods, have revealed how eukaryotic transcription is driven by a dizzying array of molecular processes that span a wide range of timescales. The overview of these timescales presented in \autoref{fig:BurstingVsMoleculesTimesScales}E show how many of these processes are significantly faster than transcriptional bursting. 

Chromatin accessibility is a central control point for regulating transcription in eukaryotes \parencite{Eck2020,Bowman2015}. DNA wrapped around nucleosome restricts transcription factor access \parencite{Kouzarides2007,Bowman2015}. Multiple studies have determined the timescales of spontaneous DNA unwrapping and rewrapping to be around 0.01-5~s \parencite{Tomschik2005,Li2005,Kassabov2003}. 
While unwrapping and rewrapping are probably too fast to directly lead to long transcriptional bursts, DNA unwrapping might represent a ``foothold'' by which factors transiently bind DNA and enact larger-scale, sustained chromatin modifications \parencite{Bowman2015}. 

Interestingly, nucleosome turnover occurs over a longer timescale compatible with bursting, with multiple studies suggesting timescales of several minutes to hours \parencite{Deal2010,Ghuysen2007,Kimura2001,Misteli2000,Waterborg1993}. Recent genome-wide studies have measured average nucleosome turnover time to be approximately 1 hour in the fly and in yeast \parencite{Deal2010,Ghuysen2007}.
Further, histone modifications may modulate nucleosomal occupancy \parencite{Bowman2015,Ludwig2019,Lawrence2016}, and the half-life as well as addition of these modifications can also span a broad range of timescales compatible with bursting, from several minutes to days \parencite{Braun2017,Bintu2016a,Hathaway2012,Zee2010,katan2002dynamics,Chestier1979,Jackson1975}.

Once the chromatin is open, enhancers, DNA stretches containing transcription factor binding sites and capable of contacting promoters to control gene expression, become accessible. Transcription factor binding recruits co-factors and general transcription factors to the promoter, triggering the biochemical cascade that ultimately initiates transcription \parencite{Coulon2013}.
While the resulting bursts of RNAP initiation last from a few minutes to hours (\autoref{fig:BurstingVsMoleculesTimesScales}A-D), single-molecule live imaging has shown that transcription factor binding is a highly transient process, with residence times of 0.5-15~s \parencite{Mir2018b,Mir2017,Dufourt2018a,Chen2014,Morisaki2014,Gebhardt2013,Mazza2012,Speil2011,Zhang2016}.
The vast majority of transcription factors bind DNA for seconds, but it is worth noting that some transcription factors and chromatin proteins can bind DNA for minutes
 \parencite{Teves2018,Hansen2017a}.
 
However, the binding of transcription factors, the general transcriptional machinery, and RNAP to the DNA might be more complex than the simple cartoon picture of individual molecules engaging and disengaging from the DNA. For example, recent experiments have revealed that both mediator and RNAP form transient clusters with relatively short lifetimes in mammalian nuclei of 5-13~s, 10~s, respectively \parencite{Cho2018,Cho2016,Chen2016,Cisse2013}. In addition, it is demonstrated that transcription factors can also form clusters \textit{in vivo} \parencite{Mir2018b,Mir2017}.
However, how these cluster dynamics relate to transcriptional activity remains unclear.

Further, enhancers and promoters are often separated by kbp to even Mbp. The mechanism by which enhancers find their target loci from such a large distance, and how this contact triggers transcription, remain uncertain and are reviewed in \parencite{Furlong2018}. \textit{In vivo} measurements of enhancer-promoter separation in the \textit{Drosophila} embryo have shown that this distance fluctuates with a timescale of tens of seconds to several minutes \parencite{Heist2019,Lim2018,Chen2018}
---timescales strikingly similar to those of bursting. However, recent work has cast doubt on the simple ``lock and key'' model of enhancer association (stable, direct contact between enhancers and promoters triggers transcription), suggesting instead that enhancers may activate cognate loci from afar and, in some cases, may activate multiple target loci simultaneously \parencite{Chen2018,Fukaya2016c,Lim2018,Furlong2018,Gu2018,Benabdallah2019,Schoenfelder2019a}. Many important questions remain about the nature of enhancer-driven activation and it remains to be seen whether enhancer association dynamics are generic aspect of eukaryotic transcriptional regulation, or whether they only pertain to a subset of organisms and genes.

A single transcriptional burst generally consists of multiple RNAP initiation events ($\sim$10-100 at a rate of 1/6-1/3~s when the promoter is ON in \textit{Drosophila}, for instance) \parencite{Lammers2020,Tantale2016,Bothma2014,Garcia2013}. The transcriptional bursting cycle thus encompasses a smaller, faster biochemical cycle in which RNAP molecules are repeatedly loaded and released by the general transcription machinery. One interesting hypothesis for the molecular origin of transcriptional bursting is that the OFF state between bursts is enacted by an RNAP molecule that becomes paused at the promoter, effectively creating a traffic jam \parencite{Jonkers2015}. Live imaging and genome-wide studies have shown that RNAP pausing before initiation is common in eukaryotes \parencite{Bartman2019a,Jonkers2015,Gaertner2014,Core2008} and that its half-life of up to 20 min can be consistent with transcriptional bursting \parencite{Steurer2018,Henriques2018,Shao2017,Krebs2017,Buckley2014,Jonkers2014,Henriques2013,Darzacq2007}.

Although the dynamics of some of the molecular processes outlined above are compatible with the long timescales of transcriptional bursting, we still lack a holistic picture of how these kinetics are integrated to realize transcriptional bursts and, ultimately, to facilitate the regulation of gene expression by transcription factors.

We must also acknowledge that we still lie at the very beginning of a reckoning with the dynamics of transcriptional processes as measurements for some molecular processes results in a range of timescales that are difficult to reconcile. In particular, we still lack solid dynamic measurements regarding the assembly of the transcription preinitiation complex. Yet, perhaps more egregious than the lack of any individual dynamical measurement is the lack of a comprehensive, quantitative, and predictive understanding of how these molecular processes interact with one another in time and space to give rise to transcriptional bursting.

\begin{scriptsize}
\renewcommand*{\arraystretch}{1.75}

\begin{longtable}[h!]{m{3.5cm} m{2.5cm} m{3cm} m{3cm} m{2cm}}
\caption{{\bf Literature summary of transcriptional bursting.} We attempted to summarize the duration of a single transcriptional burst from various organisms and genes. In the cases where the single-cell data are not available, such as in data stemming from smFISH experiments, we used population averaged $T_{ON}$ and/or $T_{OFF}$ values instead to give a sense on the timescales.}
\label{tab:bursting}\\
\hline
System & Method & Bursting Timescale & Reference \\
\hline
\multicolumn{2}{l}{\textbf{Bacteria}}\\
\hline
\rowcolor[gray]{.95}
\textit{in vitro} & single-molecule assay&  5-8\,minutes &  \parencite{Chong2014a}\\
\rowcolor[gray]{.85}
\textit{Tet} system & MS2 &  $T_{ON} \approx$ 6\,minutes, $T_{OFF} \approx$ 37 \,minutes & \parencite{Golding2005}  \\
\hline

\multicolumn{2}{l}{\textbf{Fruit fly embryo}}\\
\hline
\rowcolor[gray]{.95}
\textit{even-skipped} stripe 2 & MS2 & few \,minutes & \parencite{Lammers2020,Bothma2014}\\
\rowcolor[gray]{.85}
\textit{even-skipped} & MS2 & few \,minutes &  \parencite{Berrocal2018a}\\
\rowcolor[gray]{.95}
\textit{Notch} signaling & MS2 & 5-20 \,minutes  & \parencite{Falo-Sanjuan2019}\\
\rowcolor[gray]{.85}
\textit{snail, Kr\"upple} & MS2 & 5\,minutes  &  \parencite{Fukaya2016c}\\
\rowcolor[gray]{.95}
gap genes: \textit{hunchback, giant, Kr\"upple, knirps} & smFISH & $T_{ON} \approx$ 3 \,minutes, $T_{OFF} \approx$ 6 \,minutes & \parencite{Zoller2018}\\
\rowcolor[gray]{.85}
\textit{hunchback} & MS2 & few \,minutes  & \parencite{Desponds2016e}\\
\rowcolor[gray]{.95}
\textit{even-skipped} stripe 2 & MS2 & few \,minutes  & \parencite{Bothma2014}\\
\hline
\multicolumn{2}{l}{\textbf{Nematode}}\\
\hline
\rowcolor[gray]{.95}
\textit{Notch} signaling & MS2 & 10-70 \,minutes &  \parencite{Lee2019}\\
\hline
\multicolumn{2}{l}{\textbf{Human, Mouse}}\\
\hline
\rowcolor[gray]{.95}
\textit{TGF-}$\beta$ signaling & luciferase assay & few hours &  \parencite{Molina2013}\\
\rowcolor[gray]{.85}
\textit{TFF-1} signaling & MS2 & few hours & \parencite{Rodriguez2019}\\
\rowcolor[gray]{.85}
liver genes & smFISH & $T_{ON} \approx$ 30 \,minutes - 2 hours & \parencite{Halpern2015}\\
\rowcolor[gray]{.95}
mammalian genes & luciferase assay &  few hours & \parencite{Suter2011}\\
\hline
\multicolumn{2}{l}{\textbf{Amoeba}}\\
\hline
\rowcolor[gray]{.95}
actin gene family & RNA-seq & few hours &\parencite{Tunnacliffe2018}\\
\rowcolor[gray]{.85}
actin gene family & MS2 & 10-15 \,minutes &\parencite{Corrigan2016}\\
\hline
\end{longtable}
\end{scriptsize}

\restoregeometry

\newpage
\newgeometry{textwidth=17cm}

\begin{scriptsize}
\renewcommand*{\arraystretch}{1.75}
\begin{longtable}[!htp]{ m{3.5cm}  m{3.5cm}  m{3.5cm}  m{3cm}  m{1.5cm}}
\caption{{\bf Summary of measured timescales of underlying molecular processes associated with transcription.}\protect\footnotemark}
\label{tab:timescale}\\
\hline
System &Organism &Experimental method &Timescale &Reference \footnotetext{ While the vast majority of transcription factors bind DNA for seconds, it is worth noting that some transcription factors (e.g. TATA-binding protein) and chromatin proteins (e.g. CTCF, Cohesin) can bind DNA for minutes. These outliers are not included in \autoref{fig:BurstingVsMoleculesTimesScales}.}\\
\hline
\multicolumn{2}{l}{\textbf{Nucleosomal DNA Wrapping/Unwrapping}}\\
\hline
\rowcolor[gray]{.95}
Mononucleosomes &\textit{In vitro} reconstitution &FRET &0.1-5\,s &\parencite{Tomschik2005}\\
\rowcolor[gray]{.85}
Mononucleosomes &\textit{In vitro} reconstitution &FRET &10-250\,ms &\parencite{Li2005}\\
\rowcolor[gray]{.95}
Mononucleosomes &\textit{In vitro} reconstitution &Photochemical crosslinking &$<$1\,s &\parencite{Kassabov2003}\\
\hline
\multicolumn{2}{l}{\textbf{Nucleosome Turnover}}\\
\hline
\rowcolor[gray]{.95}
Histone H3.3 &Fruit fly cell  &Genome-wide profiling &1-1.5\,h &\parencite{Deal2010}\\
\rowcolor[gray]{.85}
Histone H3 &Yeast &Genomic tiling arrays &$\sim$1\,h &\parencite{Ghuysen2007}\\
\rowcolor[gray]{.95}
Histone H2B, H3, and H4 tagged with GFP &Human cell &FRAP &several minutes &\parencite{Kimura2001}\\
\rowcolor[gray]{.85}
Histone H1 tagged with GFP &Human cell &FRAP &several minutes &\parencite{Misteli2000}\\
\rowcolor[gray]{.95}
Histone H3 &Plant cell  &Isotope labeling &several hours &\parencite{Waterborg1993}\\
\hline
\multicolumn{2}{l}{\textbf{Histone Modification}}\\
\hline
\rowcolor[gray]{.95}
dCas9 inducible recruitment &Mammalian cell &Single-cell imaging &several hours to days &\parencite{Braun2017}\\
\rowcolor[gray]{.85}
rTetR inducible recruitment &Mammalian cell &Single-cell imaging &several hours to days &\parencite{Bintu2016a}\\
\rowcolor[gray]{.95}
Chemical-mediated recruitment &Mammalian cell &Chromatin \textit{in vivo} assay &several days &\parencite{Hathaway2012}\\
\rowcolor[gray]{.85}
Histone H3 &Human cell &Liquid chromatography, mass spectrometer &several hours to days (half-maximal time of methylation) &\parencite{Zee2010}\\
\rowcolor[gray]{.95}
Targeted recruitment &Yeast &ChIP &5-8min (reversal of targeted deacetylation) 1.5\,min(reversal of targeted acetylation) &\parencite{katan2002dynamics} \\
\rowcolor[gray]{.85}
Histone H2a, H2b, H3, and H4 &Mammalian cell &Isotope labeling &$<$15\,min (acetylation half-life) &\parencite{Chestier1979}\\
\rowcolor[gray]{.95}
Histone H2, H2a and H2b &Mammalian cell &Isotope labeling &$\sim$3\,min (acetylation half-life) & \parencite{Jackson1975}\\
\hline
\newpage
\hline
\multicolumn{2}{l}{\textbf{Transcription Factor Binding}}\\
\hline
\rowcolor[gray]{.95}
Bicoid &Fruit fly embryo &SMT &$\sim$2\,s &\parencite{Mir2018b}  \\
\rowcolor[gray]{.85} 
Bicoid &Fruit fly embryo &SMT &$\sim$1\,s &\parencite{Mir2017}\\
\rowcolor[gray]{.95}
Zelda &Fruit fly embryo&SMT &$\sim$5\,s &\parencite{Mir2018b} \\
\rowcolor[gray]{.85}
Zelda & Fruit fly embryo&FRAP, FCS &$\sim$2-3\,s &\parencite{Dufourt2018a} \\
\rowcolor[gray]{.95}
Sox2 &Mammalian cell &SMT &$\sim$12-15\,s &\parencite{Chen2014}\\
\rowcolor[gray]{.85}
p53 &Mammalian cell &SMT &$\sim$3.5\,s &\parencite{Morisaki2014}\\
\rowcolor[gray]{.95}
p53 &Mammalian cell &SMT, FRAP, FCS &$\sim$1.8\,s &\parencite{Mazza2012}\\
\rowcolor[gray]{.85}
Glucocorticoid receptor &Mammalian cell &SMT &$\sim$8.1\,s &\parencite{Morisaki2014} \\
\rowcolor[gray]{.95}
Glucocorticoid receptor &Mammalian cell &SMT &$\sim$1.45\,s &\parencite{Gebhardt2013}\\
\rowcolor[gray]{.85}
STAT1 &Mammalian cell &SMT &$\sim$0.5\,s &\parencite{Speil2011} \\
\rowcolor[gray]{.95}
TFIIB &\textit{In vitro} reconstitution &SMT &$\sim$1.5\,s &\parencite{Zhang2016}\\
\rowcolor[gray]{.85}
TATA-binding protein & Mammalian cell & SMT & 1.5-2\,min & \parencite{Teves2018}\\

\hline
\multicolumn{2}{l}{\textbf{Chromatin Protein Binding}}\\
\hline
\rowcolor[gray]{.95}
CTCF & Mammalian cell & SMT & $\sim$1-2\,min & \parencite{Hansen2017a}\\
\rowcolor[gray]{.85}
Cohesin & Mammalian cell & SMT & $\sim$22\,min & \parencite{Hansen2017a}\\
\hline
\multicolumn{2}{l}{\textbf{RNAP Cluster Dynamics}}\\
\hline
\rowcolor[gray]{.95}
RNAP tagged with Dendra2 &Mammalian cell &tcPALM &$\sim$12.9\,s (with small fraction of stable clusters) &\parencite{Cho2018}\\
\rowcolor[gray]{.85}
RNAP tagged with Dendra2 &Mammalian cell &tcPALM &$\sim$8.1\,s &\parencite{Cho2016}\\
\rowcolor[gray]{.95}
RNAP tagged with Dendra2 &Human cell  &Bayesian nanoscopy &several seconds &\parencite{Chen2016}\\
\rowcolor[gray]{.85}
RNAP tagged with Dendra2 &Human cell &tcPALM &$\sim$5.1\,s &\parencite{Cisse2013}\\
\hline
\multicolumn{2}{l}{\textbf{Mediator Cluster Dynamics}}\\
\hline
\rowcolor[gray]{.95}
Mediator tagged with Dendra2 &Mammalian cell &tcPALM &$\sim$11.1\,s &\parencite{Cho2018}\\
\hline
\multicolumn{2}{l}{\textbf{Enhancer-Promoter Interaction}}\\
\hline
\rowcolor[gray]{.95}
\textit{snail} shadow enhancer &Fruit fly embryo &MS2, PP7 labeling &$\sim$10-40\,s (fluctuation cycle interval) &\parencite{Heist2019}\\
\rowcolor[gray]{.85}
\textit{snail} enhancer &Fruit fly embryo &MS2, PP7 labeling &several minutes &\parencite{Lim2018}\\
\rowcolor[gray]{.95}
endogenous \textit{even-skipped} locus with homie insulator &Fruit fly embryo &MS2, PP7 labeling &several minutes &\parencite{Chen2018}\\

\hline
\multicolumn{2}{l}{\textbf{Transcription Initiation}}\\
\hline
\rowcolor[gray]{.95}
\textit{even-skipped} stripe 2 enhancer &Fruit fly embryo &MS2 labeling &$\sim$3\,s (promoter ON) &\parencite{Lammers2020}\\
\rowcolor[gray]{.85}
HIV-1 promoter &Mammalian cell &MS2 labeling &$\sim$4.1\,s (promoter ON) &\parencite{Tantale2016}\\
\rowcolor[gray]{.95}
\textit{hb} P2 enhancer &Fruit fly embryo &MS2 labeling &$\sim$6\,s &\parencite{Garcia2013}\\
\hline
\multicolumn{2}{l}{\textbf{Promoter-Proximal Pausing}}\\
\hline
\rowcolor[gray]{.95}
RNAP tagged with GFP &Human cell &FRAP &$\sim$40\,s &\parencite{Steurer2018}\\
\rowcolor[gray]{.85}
RNAP (genome-wide) &Fruit fly cell &RNA sequencing &$\sim$2-20\,min &\parencite{Henriques2018}\\
\rowcolor[gray]{.95}
RNAP (genome-wide) &Fruit fly cell &ChIP-nexus &$\sim$5-20\,min &\parencite{Shao2017}\\
\rowcolor[gray]{.85}
RNAP (genome-wide) &Fruit fly cell &Genome-wide footprinting &$\sim$2.5-20\,min &\parencite{Krebs2017}\\
\rowcolor[gray]{.95}
RNAP tagged with GFP & Fruit fly salivary glands & Single-cell imaging &$\sim$5\,min &\parencite{Buckley2014} \\
\rowcolor[gray]{.85}
RNAP (genome-wide) &Mammalian cell &GRO-seq &$\sim$6.9\,min (average) &\parencite{Jonkers2014}\\
\rowcolor[gray]{.95}
RNAP (genome-wide) &Fruit fly cell &scRNA-seq &15-20\,min (at genes with low activity) &\parencite{Henriques2013}\\
\rowcolor[gray]{.85}
LacO-tagged minimal CMV promoter &Human cell &MS2 labeling, FRAP &$\sim$4\,min &\parencite{Darzacq2007} \\
\hline
\end{longtable}

\indent SMT: Single-molecule tracking\\
\indent FRAP: Fluorescence recovery after photobleaching\\
\indent FCS: Fluorescence correlation spectroscopy\\
\indent FRET: Fluorescence resonance energy transfer\\
\indent ChIP: Chromatin immunoprecipitation\\
\indent PALM: Photo-activated localization microscopy
\end{scriptsize}

\bigskip

\restoregeometry

\section{Two-state model calculations}\label{app:two_state_model}

As noted in the main text, the average initiation rate is equal to $r$ times the average fraction of time the promoter spends in this ON state $p_{on}$,
\begin{equation} 
    \Big \langle \mbox{initiation rate} \Big \rangle  = r \, p_{on}.
\end{equation}
To predict the effect of bursting on transcription initiation, it is necessary to determine how $p_{on}$ depends on the bursting parameters. In the mathematical realization of the two-state model shown in Figure~\ref{fig:Bursting_features}A, the temporal evolution of $p_{off}$, the average probability of being in the OFF state, and of $p_{on}$ is given by
\begin{equation}\label{eq:poff}
    {d p_{off} \over dt} = - k_{on} \, p_{off} + k_{off} \, p_{on},
\end{equation}
and
\begin{equation}\label{eq:pon}
    {d p_{on} \over dt} = k_{on} \, p_{off} - k_{off} \, p_{on}.
\end{equation}
To solve these equations, we make the simplifying assumption that our system is in steady state such that these average probabilities of finding the system on the ON and OFF states
are constant in time. In this scenario, we can set the rates $d p_{off} / dt$ and $d p_{on} / dt$ to zero. We then solve for $\koff$ in terms of $\kon$ resulting in
\begin{equation}
    k_{off} = \frac{k_{on} \, p_{off}}{p_{on}}.
\end{equation}
Plugging in the normalization condition $p_{on} + p_{off}=1$\ gives us
\begin{equation}
    k_{off} = \frac{k_{on} \, (1-p_{on})}{p_{on}},
\end{equation}
which can be solved in terms of $k_{on}$, $k_{off}$, resulting in
\begin{equation}\label{eq:pon_sol}
    p_{on} = {k_{on} \over k_{on} + k_{off}}.
\end{equation}

\section{Molecular model calculations} \label{app:molecular_model_calc}
\label{app:molecular_model_calc_old}
Here we provide a brief overview of the calculations relating to the three theoretical models of transcription presented in Section~\ref{sec:bridge_gap}: the independent binding model (\autoref{fig:2models}C), the cooperative binding model (\autoref{fig:2models}E) and the rate-limiting step model (\autoref{fig:2models}G). We also provide resources relating to the calculation of first-passage time distributions discussed in \autoref{sec:pt_times}.

\subsection{Stochastic simulations}
\label{app:stoch_sim}

We make heavy use of stochastic simulations throughout this work. A custom-written implementation of the Gillespie Algorithm \parencite{Gillespie1977} was used to simulate trajectories for the various models discussed in the main text. These simulated trajectories were used to generate the activity trace plots in \autoref{fig:2models}D, F, and G, as well as the first-passage time distributions in \autoref{fig:first_passage_times}B-D. All code related to this project (including the Gillespie Algorithm implementation for stochastic activity trace generation) can be accessed on  GitHub \parencite{Lammers2020c}.

 \subsection{Independent binding model}
\label{app:independent_binding_calc}
All calculations in this section pertain to the independent binding model presented in \autoref{fig:2models}C.

\subsubsection{Calculating state probabilities}\label{app:StateProbIndependent}
Calculating the probability of each activity state is central to determining a system's overall transcriptional behavior. Because our mathematical model is a linear chain with no cycles (see \autoref{fig:2models}B), we can make progress towards calculating the steady state probabilities, $p_i$, by imposing detailed balance, which gives
\begin{equation}
    p_n \kp(n) = p_{n+1} \km(n+1) \label{eq:detailed_balance1},
\end{equation}
where $\kp$ and $\km$ are the effective rates of adding and subtracting a single activator molecule that we define in \autoref{fig:2models}B. Plugging in \autoref{eq:ind_on_rates_main} and \autoref{eq:ind_off_rates_main} from the main text results in
\begin{equation}
    p_n (N-n) k_{n,n+1} = p_{n+1}  (n+1) k_{n+1,n} \label{eq:detailed_balance2},
\end{equation}
where, the rates $k_{n,n+1}$ and $k_{n+1,n}$ are the microscopic binding and unbinding rates defined in \autoref{fig:2models}A, respectively. Now we make use of the fact that there are only two unique microscopic rates in the independent binding system: activator molecules bind at a rate $k_{n,n+1} = k^{b} = k^{b}_0 [A]$, with $[A]$ being the activator concentration and $k^{b}_0$ the binding rate constant, and unbind at a rate $k_{n,n-1} = k^u$. Plugging these values into \autoref{eq:detailed_balance2} and rearranging leads to
\begin{equation}
    p_{n+1}= \Big(\frac{N-n}{n+1}\Big ) \Big ( \frac{k^b}{k^u}\Big ) p_n  \label{eq:detailed_balance3}.
\end{equation}
To further simplify the expression in \autoref{eq:detailed_balance3}, we write $\frac{k^u}{k^b}$ as a dissociation constant ($K_d$), resulting in
\begin{equation}
    p_{n+1}= \Big(\frac{N-n}{n+1}\Big )  \frac{p_n}{K_d} \label{eq:detailed_balance4},
\end{equation}
which has the form of a recursive formula for calculating state probabilities from their predecessors. For instance, for the case where $n=0$ we have
\begin{equation}
    p_{1}= N \frac{p_0}{K_d}.
\end{equation}
We can extend this logic to calculate the probability of any state, $n$, as a function of $p_0$, leading to
\begin{equation}
    p_{n} = \frac{N!}{(N-n)!n!} \frac{p_0}{K_d^n} = \binom{N}{n} \frac{p_0}{K_d^n}\label{eq:detailed_balance5},
\end{equation}
where we have replaced the factorial terms with the binomial coefficient ($\binom{N}{n}$) on the far right-hand side can be thought of as accounting for the fact that a given number of activators bound, $n$, may correspond to multiple microscopic binding configurations (compare \autoref{fig:2models}A~and~B). Note that $\binom{N}{0}=1$, which means that \autoref{eq:detailed_balance5} is valid even when $n=0$. Finally, we impose the normalization condition that the sum of the state probabilities should be equal to 1, which leads to
\begin{equation}
    p_n =\frac{ p_0 \binom{N}{n} K_d^{-n}}{ p_0 \sum_{i=0}^N \binom{N}{i} K_d^{-i}} \label{eq:full_ind_probs}.
\end{equation}
Canceling out the factors of $p_0$ gives us our final expression for $p_n$, namely
\begin{equation}
    p_n =\frac{\binom{N}{n} K_d^{-n}}{\sum_{i=0}^N \binom{N}{i} K_d^{-i}} = \frac{\binom{N}{n} K_d^{-n}}{Z} \label{eq:full_ind_probs_simp},
\end{equation}
where $Z$ on the far righ-hand side indicates the sum of all state weights. Thus, given values of the rates $k^b$ and $k^u$, which define $K_d$, we can calculate the probability of the system being in each binding state $n$. This probability is shown diagrammatically in the shading of the different states in \autoref{fig:2models}C.

\subsubsection{Independent binding cannot produce bimodal transcriptional output} \label{app:independen_binding_no_bursting}
A basic requirement for bimodal transcriptionl behavior is that $p_0>p_1$ and $p_{N}>p_{N-1}$, where $N$ is the total number of binding sites. Couching this in terms of \autoref{eq:full_ind_probs_simp} leads to
\begin{equation}
    \frac{p_0}{p_1} = \frac{1}{N}K_d > 1,
\end{equation}
which simplifies to
\begin{equation}\label{eq:KdLarger}
    K_d > N
\end{equation}
for the low activity regime and
\begin{equation}
    \frac{p_N}{p_{N-1}} = \frac{1}{N}\frac{1}{K_d} > 1,
\end{equation}
leading to
\begin{equation}\label{eq:KdSmaller}
    K_d < \frac{1}{N}
\end{equation}
for the high activity regime. Since $K_d$ is set by the ratio $\frac{k^u}{k^b}$, which is constant for all states in the independent binding model, it is not possible for it to be simultaneously larger (\autoref{eq:KdLarger}) and smaller (\autoref{eq:KdSmaller}) than the number of binding sites $N$. We thus conclude that independent binding is incompatible with bimodal transcription, regardless of the number of binding sites $N$.

\subsubsection{Diffusion-limited binding}
\label{app:diffusion_kb}
In the main text we state that we set $k^b = [A]k^b_0$ to $0.5~\mbox{s}^{-1}$ for the simulations shown in \autoref{fig:2models}C. This is convenient because it leads to a model where half the available sites are bound, on average. This choice is also physically reasonable. Bicoid concentrations in the anterior region of the embryo (where {\it hunchback} is expressed) are on the order of $30~\text{molecules per}~\mu \text{m}^{3}$ \parencite{Gregor2007a}. A $k^b$ of $0.5~\mbox{s}^{-1}$ thus implies that  $k^b_0\approx0.017~\mu\text{m}^{3}\text{s}^{-1}$ per molecule. This falls below a recent estimate for the upper limit on $k^b_0$ for Bicoid binding set by diffusion of $\sim 0.022 ~\mu\text{m}^{3}\text{s}^{-1}$ per molecule \parencite{Desponds2020a}. 

We also note here that the largest binding rate in the cooperative binding model (\autoref{fig:2models}E~and~F), $k^b=58~\text{s}^{-1}$, implies a $k^b_0$ that is significantly above diffusion limit diffusion limit for Bicoid estimated in \parencite{Desponds2020a}. This high binding rate implies that cooperative binding interactions somehow facilitate the super-diffusive recruitment of additional activator molecules to the gene locus. While speculative, we note that the relatively small energies needed to realized this rapid recruitment in our model (for the plots in \autoref{fig:2models}, the cooperativity factor $\omega$ equals 6.7, which corresponds to protein-protein interactions with energies of $1.9 k_b T$), suggest that this kind of behavior is at least physically plausible. Alternatively, a significant increase in the concentration of activator molecules in the direct vicinity of the gene locus could facilitate rapid activator binding without exceeding the limits set by diffusion. Recent experiments provide evidence for this kind of local enrichment \parencite{Dufourt2018a,Mir2018b}, but it remains to be seen whether this phenomenon plays a role in facilitating gene regulation and, in particular, whether local transcription factor enrichment influences bursting. 

Finally, we also note that the rate-limiting step model (\autoref{fig:2models}G~and~H) assumes a binding rate in the ON state, $k^b_{on}=21~s^{-1}$, that likewise implies a $k^b_0$ that is above Bicoid's likely diffusion limit. This high binding rate was employed primarily for clarity of exposition, ensuring that the rate-limiting steps separated ``ON'' and ``OFF'' activity regimes that were well resolved from one another. Our principal conclusions do not depend on the precise value of $k^b_{on}$, though, naturally, inferring the durations of OFF (and ON) periods as was done for the waiting time analyses in \autoref{fig:first_passage_times} becomes more difficult. Specifically, smaller values of $k^b_{on}$ reduce the average number of molecules bound when the system is in the ON state,
\begin{equation}
     \langle n_{on} \rangle = N\frac{k_{on}^b}{k_{on}^b + k_{on}^u} \label{eq:number_bound},
\end{equation}
which leads to more overlap between the transcriptional activity corresponding to the ON and OFF states. It is plausible that a $k^b_{on}$ of this magnitude could be realized by other activator molecules that (i) diffuse faster than Bicoid, (ii) have larger binding target regions, or (iii) are expressed at higher concentrations endogenously. Alternatively, \autoref{eq:number_bound} indicates that a rate-limiting step mechanism that alters the rate of unbinding ($k^u$) when switching between ON and OFF states instead of, or in conjunction with, $k^b$ could lead to similarly well-resolved ON and OFF states to those in \autoref{fig:2models}H at much lower values of $k^b$.

\subsection{Cooperative binding} 
\label{app:cooperative_binding_appendix}
All calculations in this section pertain to the independent binding model presented in \autoref{fig:2models}E.

\subsubsection{Deriving state probabilities with cooperative binding}
In \autoref{eq:coop_binding_micro} of the main text we incorporated cooperativity to binding by adding multiplicative weights, $\omega$, giving
\begin{equation}
    k_{i,i+1} = k^b \omega^i. 
\end{equation}
This functional form follows from the assumption that each bound activator increases $k^b$ by a constant factor $\omega \geq 1$. This leads the expression for $\kp(n)$
\begin{equation}
    \kp^{coop}(n) = (N-n) \omega^n k^b  \label{eq:coop_on_rates_app},
\end{equation}
which is a nonlinear function of $n$. Now, in analogy to the calculations presented in \ref{app:StateProbIndependent}, let's re-derive our expressions for $p_n$. To start, we have
\begin{equation}
    p_{n+1}= \Big(\frac{N-n}{n+1}\Big ) \Big ( \frac{k^b}{k^u}\Big ) \omega^n  p_n  \label{eq:coop_detailed_balance3}.
\end{equation}
Again expressing $\frac{k^u}{k^b}$ as a dissociation constant ($K_d$), we obtain
\begin{equation}
    p_{n+1}= \Big(\frac{N-n}{n+1}\Big )  \frac{\omega^n p_n}{K_d} \label{eq:coop_detailed_balance4}.
\end{equation}
We can also extend this logic to calculate the probability of any state, $n$, as a function of $p_0$, leading to
\begin{equation}
    p_{n} = \frac{N!}{(N-n)!n!} \frac{\omega^{\frac{n(n-1)}{2}} p_0}{K_d^n} = \binom{N}{n} \frac{\omega^{\frac{n(n-1)}{2}} p_0}{K_d^n}\label{eq:coop_detailed_balance5}.
\end{equation}
Finally, by requiring that all state probabilities sum to one, we obtain
\begin{equation}
    p_n = \frac{\binom{N}{n} \omega^{\frac{n(n-1)}{2}}K_d^{-n}}{Z} \label{eq:full_coop_probs_simp},
\end{equation}
where $Z$ again denotes the sum of all state weights as in \autoref{eq:full_ind_probs_simp}. We have used these expressions to calculate the probability of each state shown using the shading in \autoref{fig:2models}E.

\subsubsection{Cooperativity permits bimodal expression}
Now, let's use \autoref{eq:coop_detailed_balance5} to examine how the addition of the cooperativity factor $\omega$ makes bimodal bursting possible. Recall that bimodal gene expression requires that $p_0>p_1$ and $p_{N}>p_{N-1}$. For the low activity regime, cooperativity is not relevant because there are no already bound activators, and so the form of the requirement remains the same, namely
\begin{equation}
    \frac{p_0}{p_1} = \frac{1}{N} K_d > 1.
\end{equation}
However, things change in the high activity regime. Here, we have
\begin{equation}
    \frac{p_N}{p_{N-1}} = \frac{1}{N}\frac{\omega^{N-1}}{K_d} > 1.
\end{equation}
In stark contrast to the independent binding case, we see that the addition of $\omega$ makes it possible to realize both conditions simultaneously, opening the door to bimodal burst behaviors. Specifically, bimodality demands 
\begin{equation}
    K_d > N,
\end{equation}
and 
\begin{equation}
    \omega > (N K_d)^{\frac{1}{N-1}}
\end{equation}
to be true. 

These requirements thus demonstrate that cooperativity is required to achieve bimodal bursting in the context of this binding model. Indeed, the cooperative binding system shown in \autoref{fig:2models}E~and~F meets these criteria, having $K_d=116$ and $\omega=6.7$, both of which are comfortably above the lower bounds described above for a system with six binding sites ($N=6$). 
\subsubsection{Cooperativity is necessary to simultaneously achieve kinetic trapping at both ends of the chain} 
In the main text we introduced the concept of ``kinetic trapping''; a phenomenon in which large imbalances between between $\kp(n)$ and $\km(n)$ cause a system to get trapped in high and low activity states for periods of time that far exceed the timescale of individual binding/unbinding events. Here, we show that cooperativity ($\omega>1$) is needed in order to achieve this kind of trapping at \textit{both} the high and low ends of the binding chain shown in \autoref{fig:2models}B simultaneously.

To begin, we note that the relations $\km(1)>\kp(1)$ and $\km(N-1)<\kp(N-1)$ are necessary to have traps at the low and high ends of the chain, respectively. Thus both conditions must hold simultaneously for traps to exist at the low and high ends simultaneously. We can use \autoref{eq:coop_on_rates_app} and \autoref{eq:ind_off_rates_main} to express these requirements in terms of system parameters. For the low activity regime, we have
\begin{equation}
    \frac{\km(1)}{\kp(1)} = \frac{K_d}{(N-1) \omega} > 1, \label{eq:trapping_low_bonds}
\end{equation}
and for the higher regime we obtain
\begin{equation}
    \frac{\kp(N-1)}{\km(N-1)} = \frac{\omega^{N-1}}{(N-1) K^d} > 1.
\end{equation}
We can simplify these requirements to obtain upper and lower bounds on $\omega$, namely
\begin{equation}
    \Big[K_d (N-1) \Big]^{\frac{1}{N-1}}~ < ~\omega~<~ \frac{K_d}{N-1}. \label{eq:omega_bounds_trapping}
\end{equation}
We see that \autoref{eq:omega_bounds_trapping} implies restrictions on the relationship between $K_d$ and $N$. Specifically, there must be a gap between the upper and lower bounds in \autoref{eq:omega_bounds_trapping} such that there exist viable $\omega$ values. This means that
\begin{equation}
    \Big[K_d (N-1) \Big]^{\frac{1}{N-1}}~ < ~ \frac{K_d}{N-1},
\end{equation}
must hold. Upon simplification, this gives
\begin{equation}
    (N-1)^N ~ < ~ K_d^{N-2}.
\end{equation}
\begin{equation}
     (N-1)^2 < K_d \label{eq:Kd_bounds_trapping}.
\end{equation}

\autoref{eq:Kd_bounds_trapping} tells us that the dissociation constant must be larger than one (indeed, it must be larger than 25 for an $N=6$ binding site system). This implies that the expression for the lower $\omega$ bound on the left-hand side of \autoref{eq:omega_bounds_trapping} is guaranteed to be greater than one as well because
\begin{equation}
    K_d(N-1)~ > 1,
\end{equation}
and therefore 
\begin{equation}
    1 ~ < ~ \Big[K_d (N-1) \Big]^{\frac{1}{N-1}} ~ < ~ \omega.
\end{equation}
This indicates that cooperative interactions are necessary to realize kinetic traps on both ends of the chain, though it hints at the fact that increasing the number of binding sites, $N$, should enable trapping with lower values of $\omega$. 

\subsubsection{Off rate-mediated cooperativity can also generate transcriptional bursting} \label{app:koff_cooperativity}

\begin{figure}[p]
\includegraphics[width=160mm]{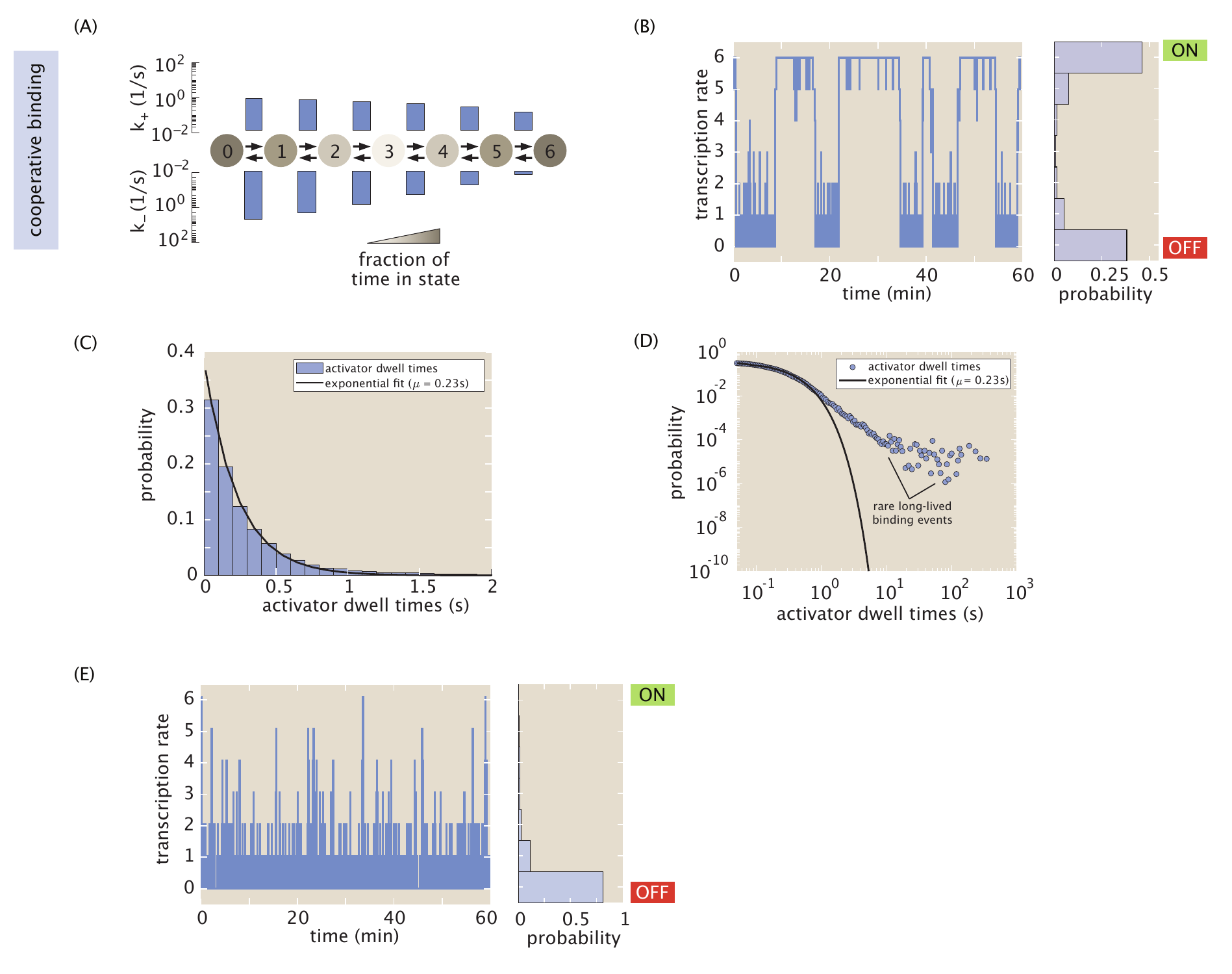}
\centering
\caption{{\bf Cooperative binding model with off rate-mediated cooperativity}. \textbf{(A)} Cooperative binding model in which stabilizing interactions between bound activators decreases the unbinding rate. \textbf{(B)} Simulation reveals that off rate-mediated cooperativity can cause the system to exhibit bimodal rates of transcription and slow fluctuations between effective ON and OFF states that are comparable to those observed for on rate-mediated cooperativity (\autoref{fig:2models}E~and~F). \textbf{(C)} The predicted distribution of unbinding times is well fit by a single exponential, despite the presence of cooperative interactions. \textbf{(D)} However, this fit masks the presence of rare long-lived binding events that are signatures of the cooperative effects that enable slow timescale bursting. These events can be seen as a significant deviation between the empirical distribution of dwell times (blue circles) from the exponential fit when we shift to looking at the data on a log-log scale. \textbf{(E)} To examine the role these long-lived events play in bursting, we conducted simulations in which all activator dwell events were capped to a maximum duration of 10 seconds; i.e. we forced all activators still bound after 10 seconds to unbind. These simulations revealed that removing long-lived binding events abolishes all burst-like activity.}
\label{fig:koff_coop}
\end{figure}

It is straight-forward to adapt our discussion of on rate-mediated cooperativity to capture the case where cooperative interactions act instead to stabilize bound factors and thereby reduce the effective off rate. Indeed, the expressions are identical save for the fact that the basal binding rate is \textit{divided} by powers of $\omega$, rather than multiplied. Namely,
\begin{equation}
    k_{i,i-1} = \frac{k^u}{\omega^i}. 
\end{equation}
 This leads the expression for $\km(n)$
\begin{equation}
    \km^{coop}(n) = n \frac{k^u}{\omega^n} \label{eq:coop_off_rates_app}.
\end{equation}
Since activator binding is assumed to be independent, the expression for $\kp(n)$ is identical to that for the independent binding model, namely,
\begin{equation}
    \kp(n) = (N-n) k_{n,n+1}  \label{eq:ind_on_rates_main}.
\end{equation}

These expressions lead to a model wherein the unbinding rate decreases significantly as more and more activators become bound (\autoref{fig:koff_coop}A). From this point forward, the expressions for state probabilities (starting with \autoref{eq:coop_detailed_balance3}) have precisely the same form as those for the on rate-mediated case. 

With the off rate-mediated model thus defined, we are in a position to once again employ stochastic simulations to explore the impact of cooperativity on transcriptional dynamics. These reveal that stabilizing cooperative interactions amongst bound activator molecules can also generate slow bimodal transcriptional fluctuations (\autoref{fig:koff_coop}B). While this is perhaps not surprising given the close mathematical parallels between the on and off rate-mediated cooperativity models, this result emphasizes the bredth of potential molecular mechanisms that could lead to transcriptional bursting.

Given these results, it is interesting to ask whether stabilizing cooperative interactions should have a clear signature in experimental measurements of individual activator residence times. Somewhat surprisingly, a distribution of single particle unbinding events generated via stochastic simulations are well described by a single exponential fit with a mean residence time of $0.23~\text{s}^{-1}$ (\autoref{fig:koff_coop}C), which is on the same scale as the empirical measurements in \parencite{Mir2018b}. Yet, this is no the full story. A close inspection of the distribution of dwell times reveals a small ($\lesssim 5 \text{\%}$)  fraction of very long lived binding events that last for 10s to 100s of seconds (\autoref{fig:koff_coop}D). 

Such events might very well be missed in an \textit{in vivo} experiment due either to how rare they are or to limitations on observation time imposed by the bleaching of fluorescent proteins. To examine the role these events play in dictating burst dynamics, we conducted simulations invoking parameters identical to those used to generate the trace shown in  \autoref{fig:koff_coop}B, with one important alteration: all activators still bound after 10 seconds were forced to unbind. This effectively capped the activator dwell time at 10 seconds, thereby abolishing all long-lived binding events. \autoref{fig:koff_coop}E shows the result of this exercise, clearly indicating, that removing rare long-lived binding events abolishes burst-like dynamics. These results thus indicate that hard-to-detect long-lived binding events could play a key role in generating slow burst dynamics, suggesting that it could be of interest to design experiments explicitly aimed at searching for long-lived binding events \textit{in vivo}. In closing, we note that similar results relating to activator dwell time distributions in the presence of cooperative interactions are discussed in \parencite{Desponds2020a}. We direct readers to this work for an excellent, detailed discussion on this topic. 

\subsection{First-passage time calculations}
\label{app:fp_calc}
In this review we used stochastic simulations (briefly outlined in \ref{app:stoch_sim}) to arrive at expectations for the form of first-passage time distributions for the cooperative binding and rate-limiting step models. All relevant scripts are available at \textcolor{blue}{\hyperlink{https://github.com/nlammers371/transcription_timescales_review.git}{GitHub}}. We also note that the functional forms for waiting time distributions can be calculated using analytical methods such as Laplace Transforms. We do not provide the details for this approach here, but point the reader to \parencite{Schnitzer1995,Bel2010}, as well as the sources cited therein, for more information.

\printbibliography

@article{Zhang2016,
author = {Zhang, Zhengjian and English, Brian P and Grimm, Jonathan B and Kazane, Stephanie A and Hu, Wenxin and Tsai, Albert and Inouye, Carla and You, Changjiang and Piehler, Jacob and Schultz, Peter G and Lavis, Luke D and Revyakin, Andrey and Tjian, Robert},
doi = {10.1101/gad.285395.116},
issn = {1549-5477},
journal = {Genes {\&} development},
number = {18},
pages = {2106--2118},
pmid = {27798851},
publisher = {Cold Spring Harbor Laboratory Press},
title = {{Rapid dynamics of general transcription factor TFIIB binding during preinitiation complex assembly revealed by single-molecule analysis.}},
url = {http://www.ncbi.nlm.nih.gov/pubmed/27798851 http://www.pubmedcentral.nih.gov/articlerender.fcgi?artid=PMC5066616},
volume = {30},
year = {2016}
}

@article{Uzman2002,
author = {Uzman, Akif},
doi = {10.1002/bmb.2002.494030059994},
issn = {14708175},
journal = {Biochemistry and Molecular Biology Education},
number = {5},
pages = {340--341},
publisher = {Wiley},
title = {{Genes and signals: Ptashne, M., Gann, A.}},
url = {http://doi.wiley.com/10.1002/bmb.2002.494030059994},
volume = {30},
year = {2002}
}

@article{Larson2011,
author = {Larson, Daniel R and Zenklusen, Daniel and Wu, Bin and Chao, Jeffrey A and Singer, Robert H},
title = {Real-Time Observation of Transcription Initiation and Elongation on an Endogenous Yeast Gene},
url = {http://science.sciencemag.org/},
volume = {332},
pages = {475--478},
year = {2011},
journal = {Science},
}

@article{Chubb2006,
author = {Chubb, Jonathan R. and Trcek, Tatjana and Shenoy, Shailesh M. and Singer, Robert H.},
doi = {10.1016/j.cub.2006.03.092},
issn = {09609822},
journal = {Current Biology},
number = {10},
pages = {1018--1025},
title = {{Transcriptional Pulsing of a Developmental Gene}},
volume = {16},
year = {2006}
}

@article{Hathaway2012,
author = {Hathaway, Nathaniel A. and Bell, Oliver and Hodges, Courtney and Miller, Erik L. and Neel, Dana S. and Crabtree, Gerald R.},
doi = {10.1016/j.cell.2012.03.052},
issn = {00928674},
journal = {Cell},
number = {7},
pages = {1447--1460},
publisher = {Elsevier Inc.},
title = {{Dynamics and memory of heterochromatin in living cells}},
url = {http://dx.doi.org/10.1016/j.cell.2012.03.052},
volume = {149},
year = {2012}
}

@article{Cho2018,
author = {Cho, Won Ki and Spille, Jan Hendrik and Hecht, Micca and Lee, Choongman and Li, Charles and Grube, Valentin and Cisse, Ibrahim I.},
doi = {10.1126/science.aar4199},
issn = {10959203},
journal = {Science},
number = {6400},
pages = {412--415},
title = {{Mediator and RNA polymerase II clusters associate in transcription-dependent condensates}},
volume = {361},
year = {2018}
}

@article{Li2005,
author = {Li, Gu and Levitus, Marcia and Bustamante, Carlos and Widom, Jonathan},
doi = {10.1038/nsmb869},
journal = {Nature Structural and Molecular Biology},
number = {1},
pages = {46--53},
title = {{Rapid spontaneous accessibility of nucleosomal DNA}},
volume = {12},
year = {2005}
}

@article{Tunnacliffe2018,
author = {Tunnacliffe, Edward and Corrigan, Adam M. and Chubb, Jonathan R.},
doi = {10.1073/pnas.1800943115},
issn = {10916490},
journal = {Proceedings of the National Academy of Sciences of the United States of America},
number = {33},
pages = {8364--8369},
publisher = {National Academy of Sciences},
title = {{Promoter-mediated diversification of transcriptional bursting dynamics following gene duplication}},
volume = {115},
year = {2018}
}

@article{Mazza2012,
author = {Mazza, Davide and Abernathy, Alice and Golob, Nicole and Morisaki, Tatsuya and McNally, James G.},
doi = {10.1093/nar/gks701},
page = {e119},
volume = {40},
issn = {03051048},
journal = {Nucleic Acids Research},
title = {{A benchmark for chromatin binding measurements in live cells}},
year = {2012}
}

@article{Ludwig2019,
author = {Ludwig, Connor H. and Bintu, Lacramioara},
doi = {10.1242/dev.170217},
issn = {14779129},
journal = {Development},
number = {12},
title = {{Mapping chromatin modifications at the single cell level}},
volume = {146},
year = {2019}
}

@article{Chen2018,
author = {Chen, Hongtao and Levo, Michal and Barinov, Lev and Fujioka, Miki and Jaynes, James B. and Gregor, Thomas},
doi = {10.1038/s41588-018-0175-z},
issn = {1061-4036},
journal = {Nature Genetics},
number = {9},
pages = {1296--1303},
publisher = {Nature Publishing Group},
title = {{Dynamic interplay between enhancer–promoter topology and gene activity}},
url = {http://www.nature.com/articles/s41588-018-0175-z},
volume = {50},
year = {2018}
}

@article{Gregor2007a,
author = {Gregor, Thomas and Tank, David W. and Wieschaus, Eric F. and Bialek, William},
doi = {10.1016/j.cell.2007.05.025},
issn = {00928674},
journal = {Cell},
number = {1},
pages = {153--164},
title = {{Probing the limits to positional information}},
url = {http://linkinghub.elsevier.com/retrieve/pii/S0092867407006629},
volume = {130},
year = {2007}
}

@article{Silk2014,
author = {Silk, Daniel and Kirk, Paul D.W. and Barnes, Chris P. and Toni, Tina and Stumpf, Michael P.H.},
doi = {10.1371/journal.pcbi.1003650},
issn = {15537358},
journal = {PLoS Computational Biology},
number = {6},
pages = {1003650},
pmid = {24922483},
publisher = {Public Library of Science},
title = {{Model Selection in Systems Biology Depends on Experimental Design}},
url = {www.bbsrc.ac.uk},
volume = {10},
year = {2014}
}

@article{Lammers2020,
author = {Lammers, Nicholas C and Galstyan, Vahe and Reimer, Armando and Medin, Sean A and Wiggins, Chris H and Garcia, Hernan G},
doi = {10.1073/pnas.1912500117},
issn = {10916490},
journal = {Proceedings of the National Academy of Sciences of the United States of America},
number = {2},
pages = {836--847},
title = {{Multimodal transcriptional control of pattern formation in embryonic development}},
volume = {117},
year = {2020}
}

@article{Suter2011,
author = {Suter, David M. and Molina, Nacho and Gatfield, David and Schneider, Kim and Schibler, Ueli and Naef, Felix},
doi = {10.1126/science.1198817},
issn = {00368075},
journal = {Science},
number = {6028},
pages = {472--474},
title = {{Mammalian genes are transcribed with widely different bursting kinetics}},
volume = {332},
year = {2011}
}

@article{Mehta2019,
journal = {arXiv},
pages = {1803.08823v3},
author = {Mehta, Pankaj and Wang, Ching-Hao and Day, Alexandre G R and Richardson, Clint and Bukov, Marin and Fisher, Charles K and Schwab, David J},
title = {{A high-bias, low-variance introduction to Machine Learning for physicists}},
year = {2019}
}

@article{Lenstra2016,
author = {Lenstra, Tineke L. and Rodriguez, Joseph and Chen, Huimin and Larson, Daniel R.},
doi = {10.1146/annurev-biophys-062215-010838},
issn = {1936-122X},
journal = {Annual Review of Biophysics},
number = {1},
pages = {25--47},
title = {{Transcription Dynamics in Living Cells}},
volume = {45},
year = {2016}
}

@article{Tantale2016,
author = {Tantale, Katjana and Mueller, Florian and Kozulic-Pirher, Alja and Lesne, Annick and Victor, Jean Marc and Robert, Marie Cecile and Capozi, Serena and Chouaib, Racha and B{\"{a}}cker, Volker and Mateos-Langerak, Julio and Darzacq, Xavier and Zimmer, Christophe and Basyuk, Eugenia and Bertrand, Edouard},
doi = {10.1038/ncomms12248},
issn = {20411723},
journal = {Nature Communications},
number = {1},
pages = {12248},
publisher = {Nature Publishing Group},
title = {{A single-molecule view of transcription reveals convoys of RNA polymerases and multi-scale bursting}},
url = {http://www.nature.com/articles/ncomms12248},
volume = {7},
year = {2016}
}

@article{Femino1998,
author = {Femino, Andrea M and Fay, Fredric S and Fogarty, Kevin and Singer, Robert H},
journal = {Science},
number = {5363},
pages = {585--590},
title = {{Visualization of Single RNA Transcripts in Situ}},
volume = {280},
year = {1998}
}

@article{Ali2020,
author = {Ali, Md Zulfikar and Choubey, Sandeep and Das, Dipjyoti and Brewster, Robert C.},
doi = {10.1016/j.bpj.2020.02.002},
issn = {15420086},
journal = {Biophysical Journal},
number = {7},
pages = {1769--1781},
pmid = {32101716},
publisher = {Biophysical Society},
title = {{Probing Mechanisms of Transcription Elongation Through Cell-to-Cell Variability of RNA Polymerase}},
url = {http://www.cell.com/article/S0006349520301193/fulltext http://www.cell.com/article/S0006349520301193/abstract https://www.cell.com/biophysj/abstract/S0006-3495(20)30119-3},
volume = {118},
year = {2020}
}

@article{Lee2019,
author = {Lee, Chang Hwan and Shin, Heaji and Kimble, Judith},
doi = {10.1016/j.devcel.2019.07.001},
issn = {18781551},
journal = {Developmental Cell},
number = {4},
pages = {426--435.e4},
publisher = {Cell Press},
title = {{Dynamics of Notch-dependent transcriptional bursting in its native context}},
volume = {50},
year = {2019}
}

@article{Garcia2013,
author = {Garcia, Hernan G. and Tikhonov, Mikhail and Lin, Albert and Gregor, Thomas},
doi = {10.1016/j.cub.2013.08.054},
issn = {18790445},
journal = {Current Biology},
number = {21},
pages = {2140--2145},
publisher = {Elsevier Ltd},
title = {{Quantitative imaging of transcription in living Drosophila embryos links polymerase activity to patterning.}},
url = {http://dx.doi.org/10.1016/j.cub.2013.08.054},
volume = {23},
year = {2013}
}

@article{Eck2020,
author = {Eck, Elizabeth and Liu, Jonathan and Kazemzadeh-Atoufi, Maryam and Ghoreishi, Sydney and Blythe, Shelby and Garcia, Hernan G},
journal = {bioRxiv},
pages = {2020.01.27.922054},
title = {{Quantitative dissection of transcription in development yields evidence for transcription factor-driven chromatin accessibility}},
year = {2020}
}

@article{Lim2018,
author = {Lim, Bomyi and Heist, Tyler and Levine, Michael and Fukaya, Takashi},
doi = {10.1016/j.molcel.2018.02.029},
issn = {10974164},
journal = {Molecular Cell},
number = {2},
pages = {287--296.e6},
publisher = {Elsevier Inc.},
title = {{Visualization of transvection in living Drosophila embryos}},
volume = {70},
year = {2018}
}

@article{Raj2008,
author = {Raj, Arjun and van den Bogaard, Patrick and Rifkin, Scott A. and van Oudenaarden, Alexander and Tyagi, Sanjay},
doi = {10.1038/nmeth.1253},
issn = {15487091},
journal = {Nature Methods},
number = {10},
pages = {877--879},
title = {{Imaging individual mRNA molecules using multiple singly labeled probes}},
volume = {5},
year = {2008}
}

@article{Molina2013,
author = {Molina, Nacho and Suter, David M. and Cannavo, Rosamaria and Zoller, Benjamin and Gotic, Ivana and Naef, F{\'{e}}lix},
doi = {10.1073/pnas.1312310110},
issn = {00278424},
journal = {Proceedings of the National Academy of Sciences of the United States of America},
number = {51},
pages = {20563--20568},
title = {{Stimulus-induced modulation of transcriptional bursting in a single mammalian gene}},
volume = {110},
year = {2013}
}

@article{Speil2011,
author = {Speil, Jasmin and Baumgart, Eugen and Siebrasse, Jan Peter and Veith, Roman and Vinkemeier, Uwe and Kubitscheck, Ulrich},
doi = {10.1016/j.bpj.2011.10.006},
issn = {00063495},
journal = {Biophysical Journal},
number = {11},
pages = {2592--2600},
title = {{Activated STAT1 transcription factors conduct distinct saltatory movements in the cell nucleus}},
volume = {101},
year = {2011}
}

@article{Sanchez2011,
author = {Sanchez, Alvaro and Garcia, Hernan G. and Jones, Daniel and Phillips, Rob and Kondev, Jan{\'{e}}},
doi = {10.1371/journal.pcbi.1001100},
issn = {1553734X},
journal = {PLoS Computational Biology},
number = {3},
pmid = {21390269},
title = {{Effect of promoter architecture on the cell-to-cell variability in gene expression}},
volume = {7},
year = {2011}
}

@article{Lawrence2016,
author = {Lawrence, Moyra and Daujat, Sylvain and Schneider, Robert},
doi = {10.1016/j.tig.2015.10.007},
issn = {13624555},
journal = {Trends in Genetics},
number = {1},
pages = {42--56},
publisher = {Elsevier Ltd},
title = {{Lateral thinking: How Histone modifications regulate gene expression}},
url = {http://dx.doi.org/10.1016/j.tig.2015.10.007},
volume = {32},
year = {2016}
}

@article{Lionnet2011,
author = {Lionnet, Timoth{\'{e}}e and Czaplinski, Kevin and Darzacq, Xavier and Shav-Tal, Yaron and Wells, Amber L. and Chao, Jeffrey A. and Park, Hye Yoon and {De Turris}, Valeria and Lopez-Jones, Melissa and Singer, Robert H.},
doi = {10.1038/nmeth.1551},
issn = {15487091},
journal = {Nature Methods},
number = {2},
pages = {165--170},
pmid = {21240280},
title = {{A transgenic mouse for in vivo detection of endogenous labeled mRNA}},
volume = {8},
year = {2011}
}

@article{Phillips2019,
author = {Phillips, Rob and Belliveau, Nathan M and Chure, Griffin and Garcia, Hernan G and Razo-Mejia, Manuel and Scholes, Clarissa},
doi = {10.1146/annurev-biophys-052118},
journal = {Annu. Rev. Biophys},
pages = {121--163},
title = {{Figure 1 Theory Meets Figure 2 Experiments in the Study of Gene Expression}},
url = {https://doi.org/10.1146/annurev-biophys-052118-},
volume = {48},
year = {2019}
}

@article{Zoller2015,
author = {Zoller, Benjamin and Nicolas, Damien and Molina, Nacho and Naef, Felix},
doi = {10.15252/msb},
journal = {Mol Syst Biol},
title = {{Structure of silent transcription intervals and noise characteristics of mammalian genes}},
volume = {11},
year = {2015}
}

@article{Donovan2019,
author = {Donovan, Benjamin T and Huynh, Anh and Ball, David A and Patel, Heta P and Poirier, Michael G and Larson, Daniel R and Ferguson, Matthew L and Lenstra, Tineke L},
doi = {10.15252/embj.2018100809},
issn = {0261-4189},
journal = {The EMBO Journal},
pmid = {31101674},
title = {{Live‐cell imaging reveals the interplay between transcription factors, nucleosomes, and bursting}},
year = {2019}
}

@article{Rosenfeld2005,
author = {Rosenfeld, Nitzan and Young, Jonathan W. and Alon, Uri and Swain, Peter S. and Elowitz, Michael B.},
doi = {10.1126/science.1106914},
issn = {00368075},
journal = {Science},
number = {5717},
pages = {1962--1965},
pmid = {15790856},
publisher = {Science},
title = {{Gene regulation at the single-cell level}},
url = {https://pubmed.ncbi.nlm.nih.gov/15790856/},
volume = {307},
year = {2005}
}

@article{Heist2019,
author = {Heist, Tyler and Fukaya, Takashi and Levine, Michael},
doi = {10.1073/pnas.1908962116},
issn = {10916490},
journal = {Proceedings of the National Academy of Sciences of the United States of America},
number = {30},
pages = {15062--15067},
title = {{Large distances separate coregulated genes in living Drosophila embryos}},
volume = {116},
year = {2019}
}

@article{Zoller2018,
author = {Zoller, B and Little, S C and Gregor, T},
doi = {10.1016/j.cell.2018.09.056},
issn = {1097-4172 (Electronic) 0092-8674 (Linking)},
journal = {Cell},
number = {3},
pages = {835--847.e25},
title = {{Diverse spatial expression patterns emerge from unified kinetics of transcriptional bursting}},
type = {Journal Article},
volume = {175},
year = {2018}
}

@article{Coulon2013,
author = {Coulon, Antoine and Chow, Carson C. and Singer, Robert H. and Larson, Daniel R.},
doi = {10.1038/nrg3484},
issn = {14710056},
journal = {Nature Reviews Genetics},
number = {8},
pages = {572--584},
publisher = {Nature Publishing Group},
title = {{Eukaryotic transcriptional dynamics: From single molecules to cell populations}},
volume = {14},
year = {2013}
}

@article{Schnitzer1995,
author = {Schnitzer, M J and Block, S M},
journal = {Cold Spring Harbor Symposia on Quantitative Biology},
issn = {00917451},
pages = {793--802},
pmid = {8824454},
title = {{Statistical kinetics of processive enzymes}},
volume = {60},
year = {1995}
}

@article{Waterborg1993,
author = {Waterborg, J. H.},
issn = {00219258},
journal = {Journal of Biological Chemistry},
number = {7},
pages = {4912--4917},
title = {{Histone synthesis and turnover in alfalfa. Fast loss of highly acetylated replacement histone variant H3.2}},
volume = {268},
year = {1993}
}

@misc{Lammers2020c,
author = {Lammers, Nicholas C. and Kim, Yang Joon and Zhao, Jiaxi and Garcia, Hernan G.},
publisher = {Github},
title = {{Code from ``A matter of time: Using dynamics and theory to uncover mechanisms of transcriptional bursting''}},
url = {https://github.com/GarciaLab/TranscriptionalTimescalesReview.git},
year = {2020}
}

@article{Devilbiss2017,
author = {Devilbiss, Frank and Ramkrishna, Doraiswami},
doi = {10.1109/JPROC.2016.2560121},
isbn = {0000000000},
issn = {15582256},
journal = {Proceedings of the IEEE},
number = {2},
pages = {330--339},
publisher = {IEEE},
title = {{Addressing the Need for a Model Selection Framework in Systems Biology Using Information Theory}},
volume = {105},
year = {2017}
}

@article{Chong2014a,
author = {Chong, Shasha and Chen, Chongyi and Ge, Hao and Xie, X. Sunney},
doi = {10.1016/j.cell.2014.05.038},
issn = {10974172},
journal = {Cell},
number = {2},
pages = {314--326},
publisher = {Cell Press},
title = {{Mechanism of transcriptional bursting in bacteria}},
volume = {158},
year = {2014}
}

@article{Henriques2018,
author = {Henriques, Telmo and Scruggs, Benjamin S. and Inouye, Michiko O. and Muse, Ginger W. and Williams, Lucy H. and Burkholder, Adam B. and Lavender, Christopher A. and Fargo, David C. and Adelman, Karen},
doi = {10.1101/gad.309351.117},
issn = {15495477},
journal = {Genes \& Development},
number = {1},
pages = {26--41},
pmid = {29378787},
title = {{Widespread transcriptional pausing and elongation control at enhancers}},
volume = {32},
year = {2018}
}

@article{Serov2017,
journal = {arXiv},
pages = {1701.06079},
author = {Serov, Alexander S. and Levine, Alexander J. and Mani, Madhav},
title = {{Abortive Initiation as a Bottleneck for Transcription in the Early Drosophila Embryo}},
url = {http://arxiv.org/abs/1701.06079},
year = {2017}
}

@article{Hansen2017a,
author = {Hansen, Anders S. and Pustova, Iryna and Cattoglio, Claudia and Tjian, Robert and Darzacq, Xavier},
doi = {10.7554/eLife.25776},
issn = {2050084X},
journal = {eLife},
pages = {e25776},
pmid = {28467304},
title = {{CTCF and cohesin regulate chromatin loop stability with distinct dynamics}},
volume = {6},
year = {2017}
}

@article{Eldar2010,
author = {Eldar, Avigdor and Elowitz, Michael B.},
journal = {Nature},
doi = {10.1038/nature09326},
issn = {14764687},
number = {7312},
pages = {167--173},
pmid = {20829787},
publisher = {Nature Publishing Group},
title = {{Functional roles for noise in genetic circuits}},
volume = {467},
year = {2010}
}

@article{Munsky2015,
author = {Munsky, Brian and Fox, Zachary and Neuert, Gregor},
journal = {Methods},
doi = {10.1016/j.ymeth.2015.06.009},
issn = {10959130},
pages = {12--21},
publisher = {Academic Press Inc.},
title = {{Integrating single-molecule experiments and discrete stochastic models to understand heterogeneous gene transcription dynamics}},
volume = {85},
year = {2015}
}

@article{Steurer2018,
author = {Steurer, Barbara and Janssens, Roel C. and Geverts, Bart and Geijer, Marit E. and Wienholz, Franziska and Theil, Arjan F. and Chang, Jiang and Dealy, Shannon and Pothof, Joris and {Van Cappellen}, Wiggert A. and Houtsmuller, Adriaan B. and Marteijn, Jurgen A.},
doi = {10.1073/pnas.1717920115},
issn = {10916490},
journal = {Proceedings of the National Academy of Sciences of the United States of America},
number = {19},
pages = {E4368--E4376},
pmid = {29632207},
title = {{Live-cell analysis of endogenous GFP-RPB1 uncovers rapid turnover of initiating and promoter-paused RNA Polymerase II}},
volume = {115},
year = {2018}
}

@article{Jonkers2014,
author = {Jonkers, Iris and Kwak, Hojoong and Lis, John T.},
doi = {10.7554/eLife.02407},
issn = {2050084X},
journal = {eLife},
number = {3},
pages = {e02407},
pmid = {24843027},
title = {{Genome-wide dynamics of Pol II elongation and its interplay with promoter proximal pausing, chromatin, and exons}},
volume = {3},
year = {2014}
}

@article{Rodriguez2019,
author = {Rodriguez, Joseph and Ren, Gang and Day, Christopher R. and Zhao, Keji and Chow, Carson C. and Larson, Daniel R.},
doi = {10.1016/j.cell.2018.11.026},
issn = {10974172},
journal = {Cell},
number = {1-2},
pages = {213--226.e18},
pmid = {30554876},
publisher = {Cell Press},
title = {{Intrinsic Dynamics of a human gene reveal the basis of expression heterogeneity}},
volume = {176},
year = {2019}
}

@article{Tomschik2005,
author = {Tomschik, Miroslav and Zheng, Haocheng and {Van Holde}, Ken and Zlatanova, Jordanka and Leuba, Sanford H.},
doi = {10.1073/pnas.0500189102},
isbn = {0500189102},
journal = {Proceedings of the National Academy of Sciences of the United States of America},
number = {9},
pages = {3278--3283},
title = {{Fast long-range, reversible conformational fluctuations in nucleosomes revealed by single-pair fluorescence resonance energy transfer}},
volume = {102},
year = {2005}
}

@article{Gillespie1977,
author = {Gillespie, Daniel T.},
journal = {Journal of Physical Chemistry},
doi = {10.1021/j100540a008},
issn = {00223654},
number = {25},
pages = {2340--2361},
publisher = {American Chemical Society},
title = {{Exact stochastic simulation of coupled chemical reactions}},
url = {https://pubs.acs.org/sharingguidelines},
volume = {81},
year = {1977}
}

@article{Luria1943,
author = {Luria, S. E. and Delbr{\"{u}}ck, M.},
journal = {Genetics},
number = {6},
title = {{Mutations of bacteria from virus sensitivity to virus resistance}},
volume = {28},
year = {1943}
}

@article{Kimura2001,
author = {Kimura, Hiroshi and Cook, Peter R.},
doi = {10.1083/jcb.153.7.1341},
journal = {Journal of Cell Biology},
number = {7},
pages = {1341--1353},
pmid = {11425866},
title = {{Kinetics of core histones in living human cells: Little exchange of H3 and H4 and some rapid exchange of H2B}},
volume = {153},
year = {2001}
}

@article{Falo-Sanjuan2019,
author = {Falo-Sanjuan, Julia and Lammers, Nicholas C and Garcia, Hernan G and Bray, Sarah J},
doi = {10.1016/j.devcel.2019.07.002},
journal = {Developmental Cell},
pages = {411--425},
title = {{Enhancer priming enables fast and sustained transcriptional responses to Notch signaling}},
url = {https://doi.org/10.1016/j.devcel.2019.07.002},
volume = {50},
year = {2019}
}

@article{Nicolas2017,
author = {Nicolas, Damien and Phillips, Nick E. and Naef, Felix},
doi = {10.1039/C7MB00154A},
issn = {1742-206X},
journal = {Molecular BioSystems},
number = {7},
pages = {1280--1290},
publisher = {The Royal Society of Chemistry},
title = {{What shapes eukaryotic transcriptional bursting?}},
url = {http://xlink.rsc.org/?DOI=C7MB00154A},
volume = {13},
year = {2017}
}

@article{Buckley2014,
author = {Buckley, Martin S. and Kwak, Hojoong and Zipfel, Warren R. and Lis, John T.},
doi = {10.1101/gad.231886.113},
issn = {08909369},
journal = {Genes \& Development},
number = {1},
pages = {14--19},
title = {{Kinetics of promoter Pol II on Hsp70 reveal stable pausing and key insights into its regulation}},
volume = {28},
year = {2014}
}

@article{Darzacq2007,
author = {Darzacq, Xavier and Shav-Tal, Yaron and {De Turris}, Valeria and Brody, Yehuda and Shenoy, Shailesh M. and Phair, Robert D. and Singer, Robert H.},
doi = {10.1038/nsmb1280},
issn = {15459993},
journal = {Nature Structural and Molecular Biology},
number = {9},
pages = {796--806},
title = {{In vivo dynamics of RNA Polymerase II transcription}},
volume = {14},
year = {2007}
}

@article{Kassabov2003,
author = {Kassabov, Stefan R. and Zhang, Bei and Persinger, Jim and Bartholomew, Blaine},
doi = {10.1016/S1097-2765(03)00039-X},
journal = {Molecular Cell},
number = {2},
pages = {391--403},
pmid = {12620227},
title = {{SWI/SNF unwraps, slides, and rewraps the nucleosome}},
volume = {11},
year = {2003}
}

@article{Henriques2013,
author = {Henriques, Telmo and Gilchrist, Daniel A. and Nechaev, Sergei and Bern, Michael and Muse, Ginger W. and Burkholder, Adam and Fargo, David C. and Adelman, Karen},
doi = {10.1016/j.molcel.2013.10.001},
issn = {10972765},
journal = {Molecular Cell},
number = {4},
pages = {517--528},
pmid = {24184211},
publisher = {Elsevier Inc.},
title = {{Stable pausing by RNA Polymerase II provides an opportunity to target and integrate regulatory signals}},
url = {http://dx.doi.org/10.1016/j.molcel.2013.10.001},
volume = {52},
year = {2013}
}

@article{Bintu2016a,
author = {Bintu, Lacramioara and Yong, John and Antebi, Yaron E. and McCue, Kayla and Kazuki, Yasuhiro and Uno, Narumi and Oshimura, Mitsuo and Elowitz, Michael B.},
doi = {10.1126/science.aab2956},
issn = {10959203},
journal = {Science},
number = {6274},
pages = {720--724},
title = {{Dynamics of epigenetic regulation at the single-cell level}},
volume = {351},
year = {2016}
}

@article{Dufourt2018a,
author = {Dufourt, Jeremy and Trullo, Antonio and Hunter, Jennifer and Fernandez, Carola and Lazaro, Jorge and Dejean, Matthieu and Morales, Lucas and Nait-Amer, Saida and Schulz, Katharine N. and Harrison, Melissa M. and Favard, Cyril and Radulescu, Ovidiu and Lagha, Mounia},
doi = {10.1038/s41467-018-07613-z},
issn = {20411723},
journal = {Nature Communications},
number = {1},
pmid = {30518940},
publisher = {Nature Publishing Group},
title = {{Temporal control of gene expression by the pioneer factor Zelda through transient interactions in hubs}},
url = {www.nature.com/naturecommunications},
volume = {9},
year = {2018}
}

@article{Suter2011c,
author = {Suter, David M. and Molina, Nacho and Naef, Felix and Schibler, Ueli},
journal = {Current Opinion in Cell Biology},
doi = {10.1016/j.ceb.2011.09.004},
issn = {09550674},
number = {6},
pages = {657--662},
title = {{Origins and consequences of transcriptional discontinuity}},
volume = {23},
year = {2011}
}

@article{Deal2010,
author = {Deal, Roger B and Henikoff, Jorja G and Henikoff, Steven},
doi = {10.1126/science.1186777},
journal = {Science},
number = {5982},
pages = {1161--1165},
pmid = {20508129},
title = {{Genome-wide kinetics of nucleosome turnover determined by metabolic labeling of histones}},
volume = {328},
year = {2010}
}

@article{Golding2005,
author = {Golding, I and Paulsson, J and Zawilski, S M and Cox, E C},
issn = {0092-8674 (Print)},
journal = {Cell},
number = {6},
pages = {1025--1036},
title = {{Real-time kinetics of gene activity in individual bacteria}},
type = {Journal Article},
volume = {123},
year = {2005}
}

@article{Boeger2015,
author = {Boeger, Hinrich and Shelansky, Robert and Patel, Heta and Brown, Christopher R.},
journal = {Genes},
doi = {10.3390/genes6030469},
issn = {20734425},
number = {3},
pages = {469--483},
publisher = {MDPI AG},
title = {{From structural variation of gene molecules to chromatin dynamics and transcriptional bursting}},
volume = {6},
year = {2015}
}

@article{Bertrand1998,
author = {Bertrand, E and Chartrand, P and Schaefer, M and Shenoy, S M and Singer, R H and Long, R M},
doi = {S1097-2765(00)80143-4 [pii]},
issn = {1097-2765 (Print) 1097-2765 (Linking)},
journal = {Mol Cell},
number = {4},
pages = {437--445},
title = {{Localization of ASH1 mRNA particles in living yeast}},
type = {Journal Article},
volume = {2},
year = {1998}
}

@article{Cai2006,
author = {Cai, Long and Friedman, Nir and Xie, X. Sunney},
doi = {10.1038/nature04599},
issn = {14764687},
journal = {Nature},
number = {7082},
pages = {358--362},
pmid = {16541077},
publisher = {Nature Publishing Group},
title = {{Stochastic protein expression in individual cells at the single molecule level}},
url = {https://www.nature.com/articles/nature04599},
volume = {440},
year = {2006}
}

@article{Chen2016,
author = {Chen, Xuanze and Wei, Mian and Zheng, M. Mocarlo and Zhao, Jiaxi and Hao, Huiwen and Chang, Lei and Xi, Peng and Sun, Yujie},
doi = {10.1021/acsnano.5b07257},
issn = {1936086X},
journal = {ACS Nano},
number = {2},
pages = {2447--2454},
title = {{Study of RNA Polymerase II clustering inside live-cell nuclei using Bayesian Nanoscopy}},
volume = {10},
year = {2016}
}

@article{Id2018,
author = {Tran, Huy and Desponds, Jonathan and {Perez Romero}, Carmina Angelica and Coppey, Mathieu and Fradin, Cecile and Dostatni, Nathalie and Walczak, Aleksandra M.},
doi = {10.1371/journal.pcbi.1006513},
isbn = {1111111111},
issn = {15537358},
journal = {PLoS Computational Biology},
number = {10},
title = {{Precision in a rush: Trade-offs between reproducibility and steepness of the hunchback expression pattern}},
url = {https://doi.org/10.1371/journal.pcbi.1006513},
volume = {14},
year = {2018},
pages = {e1006513}
}

@article{Fukaya2016c,
author = {Fukaya, Takashi and Lim, Bomyi and Levine, Michael},
doi = {10.1016/J.CELL.2016.05.025},
issn = {0092-8674},
journal = {Cell},
number = {2},
pages = {358--368},
publisher = {Cell Press},
title = {{Enhancer control of transcriptional bursting}},
url = {https://www.sciencedirect.com/science/article/pii/S0092867416305736},
volume = {166},
year = {2016}
}

@article{Berrocal2018a,
author = {Berrocal, Augusto and Lammers, Nicholas and Garcia, Hernan G. and Eisen, Michael B.},
doi = {10.1101/335901},
journal = {bioRxiv},
pages = {335901},
publisher = {Cold Spring Harbor Laboratory},
title = {{Kinetic sculpting of the seven stripes of the Drosophila even-skipped gene}},
url = {https://www.biorxiv.org/content/10.1101/335901v2.full},
year = {2020}
}

@article{Zee2010,
author = {Zee, Barry M. and Levin, Rebecca S. and Xu, Bo and LeRoy, Gary and Wingreen, Ned S. and Garcia, Benjamin A.},
doi = {10.1074/jbc.M109.063784},
isbn = {6092588854},
issn = {00219258},
journal = {Journal of Biological Chemistry},
number = {5},
pages = {3341--3350},
title = {{In vivo residue-specific histone methylation dynamics}},
volume = {285},
year = {2010}
}

@article{Corrigan2016,
author = {Corrigan, Adam M. and Tunnacliffe, Edward and Cannon, Danielle and Chubb, Jonathan R.},
doi = {10.7554/eLife.13051},
issn = {2050084X},
pages = {e13051},
journal = {eLife},
number = {FEBRUARY2016},
publisher = {eLife Sciences Publications Ltd},
title = {{A continuum model of transcriptional bursting}},
volume = {5},
year = {2016}
}

@article{Bel2010,
author = {Bel, Golan and Munsky, Brian and Nemenman, Ilya},
doi = {10.1088/1478-3975/7/1/016003},
issn = {14783975},
journal = {Physical Biology},
number = {1},
pmid = {20026876},
publisher = {Institute of Physics Publishing},
title = {{The simplicity of completion time distributions for common complex biochemical processes}},
volume = {7},
year = {2010}
}

@article{Gaertner2014,
author = {Gaertner, Bjoern and Zeitlinger, Julia},
doi = {10.1242/dev.088492},
issn = {09501991},
journal = {Development},
number = {6},
pages = {1179--1183},
pmid = {24595285},
title = {{RNA polymerase II pausing during development}},
volume = {141},
year = {2014}
}

@article{Misteli2000,
author = {Misteli, Tom and Gunjan, Akash and Hock, Robert and Bustin, Michael and Brown, David T.},
doi = {10.1038/35048610},
journal = {Nature},
number = {6814},
pages = {877--881},
title = {{Dynamic binding of histone H1 to chromatin in living cells}},
volume = {408},
year = {2000}
}

@article{Chao2008,
author = {Chao, Jeffrey A. and Patskovsky, Yury and Almo, Steven C. and Singer, Robert H.},
doi = {10.1038/nsmb1327},
issn = {15459993},
journal = {Nature Structural and Molecular Biology},
number = {1},
pages = {103--105},
title = {{Structural basis for the coevolution of a viral RNA-protein complex}},
volume = {15},
year = {2008}
}

@article{Morisaki2014,
author = {Morisaki, Tatsuya and M{\"{u}}ller, Waltraud G. and Golob, Nicole and Mazza, Davide and McNally, James G.},
doi = {10.1038/ncomms5456},
issn = {20411723},
journal = {Nature Communications},
title = {{Single-molecule analysis of transcription factor binding at transcription sites in live cells}},
volume = {5},
year = {2014}
}

@article{Raj2009,
author = {Raj, Arjun and van Oudenaarden, Alexander},
doi = {10.1146/annurev.biophys.37.032807.125928},
issn = {1936-122X},
journal = {Annual Review of Biophysics},
number = {1},
pages = {255--270},
publisher = {Annual Reviews},
title = {{Single-Molecule Approaches to Stochastic Gene Expression}},
volume = {38},
year = {2009}
}

@article{Yildiz2004,
author = {Yildiz, Ahmet and Tomishige, Michio and Vale, Ronald D. and Selvin, Paul R.},
doi = {10.1126/science.1093753},
issn = {00368075},
journal = {Science},
number = {5658},
pages = {676--678},
publisher = {American Association for the Advancement of Science},
title = {{Kinesin Walks Hand-Over-Hand}},
volume = {303},
year = {2004}
}

@article{Gu2018,
author = {Gu, Bo and Swigut, Tomek and Spencley, Andrew and Bauer, Matthew R. and Chung, Mingyu and Meyer, Tobias and Wysocka, Joanna},
doi = {10.1126/science.aao3136},
issn = {10959203},
journal = {Science},
number = {6379},
pages = {1050--1055},
title = {{Transcription-coupled changes in nuclear mobility of mammalian cis-regulatory elements}},
volume = {359},
year = {2018}
}

@article{Jonkers2015,
author = {Jonkers, Iris and Lis, John T.},
doi = {10.1038/nrm3953},
issn = {14710080},
journal = {Nature Reviews Molecular Cell Biology},
number = {3},
pages = {167--177},
publisher = {Nature Publishing Group},
title = {{Getting up to speed with transcription elongation by RNA polymerase II}},
volume = {16},
year = {2015}
}

@article{Cho2016,
author = {Cho, Won-Ki and Jayanth, Namrata and English, Brian P and Inoue, Takuma and Andrews, J Owen and Conway, William and Grimm, Jonathan B and Spille, Jan-Hendrik and Lavis, Luke D and Lionnet, Timoth{\'{e}}e and Cisse, Ibrahim I},
doi = {10.7554/eLife.13617},
issn = {2050-084X},
journal = {eLife},
pages = {e13617},
pmid = {27138339},
publisher = {eLife Sciences Publications, Ltd},
title = {{RNA Polymerase II cluster dynamics predict mRNA output in living cells.}},
url = {http://www.ncbi.nlm.nih.gov/pubmed/27138339 http://www.pubmedcentral.nih.gov/articlerender.fcgi?artid=PMC4929003},
volume = {5},
year = {2016}
}

@article{Jackson1975,
author = {Jackson, V. and Shires, A. and Chalkley, R. and Granner, D. K.},
issn = {00219258},
journal = {Journal of Biological Chemistry},
number = {13},
pages = {4856--4863},
pmid = {168194},
title = {{Studies on highly metabolically active acetylation and phosphorylation of histones}},
volume = {250},
year = {1975}
}

@article{Wang2019a,
author = {Wang, Yaolai and Ni, Tengfei and Wang, Wei and Liu, Feng},
doi = {10.1111/brv.12452},
issn = {1469185X},
journal = {Biological Reviews},
number = {1},
pages = {248--258},
publisher = {Blackwell Publishing Ltd},
title = {{Gene transcription in bursting: a unified mode for realizing accuracy and stochasticity}},
volume = {94},
year = {2019}
}

@article{Marzen2013,
author = {Marzen, Sarah and Garcia, Hernan G. and Phillips, Rob},
journal = {Journal of Molecular Biology},
doi = {10.1016/j.jmb.2013.03.013},
issn = {10898638},
number = {9},
pages = {1433--1460},
pmid = {23499654},
publisher = {Academic Press},
title = {{Statistical mechanics of Monod-Wyman-Changeux (MWC) models}},
url = {/pmc/articles/PMC3786005/?report=abstract https://www.ncbi.nlm.nih.gov/pmc/articles/PMC3786005/},
volume = {425},
year = {2013}
}

@article{katan2002dynamics,
author = {Katan-Khaykovich, Yael and Struhl, Kevin},
doi = {10.1101/gad.967302},
issn = {08909369},
journal = {Genes and Development},
number = {6},
pages = {743--752},
publisher = {Cold Spring Harbor Lab},
title = {{Dynamics of global histone acetylation and deacetylation in vivo: Rapid restoration of normal histone acetylation status upon removal of activators and repressors}},
volume = {16},
year = {2002}
}

@article{Chestier1979,
author = {Chestier, A. and Yaniv, M.},
doi = {10.1073/pnas.76.1.46},
issn = {00278424},
journal = {Proceedings of the National Academy of Sciences of the United States of America},
number = {1},
pages = {46--50},
title = {{Rapid turnover of acetyl groups in the four core histones of simian virus 40 minichromosomes}},
volume = {76},
year = {1979}
}

@book{Sivia2006,
author = {Sivia, D S and Skilling, J},
edition = {2nd},
publisher = {Oxford University Press},
series = {Oxford Science Publications},
title = {{Data Analysis - A Bayesian Tutorial}},
year = {2006}
}

@article{Desponds2020a,
author = {Desponds, Jonathan and Vergassola, Massimo and Walczak, Aleksandra M},
doi = {10.7554/elife.49758},
issn = {2050084X},
journal = {eLife},
pages = {e49758},
pmid = {32723476},
publisher = {eLife Sciences Publications, Ltd},
title = {{A mechanism for hunchback promoters to readout morphogenetic positional information in less than a minute}},
volume = {9},
year = {2020}
}

@article{Schoenfelder2019a,
author = {Schoenfelder, Stefan and Fraser, Peter},
doi = {10.1038/s41576-019-0128-0},
issn = {14710064},
journal = {Nature Reviews Genetics},
number = {8},
pages = {437--455},
pmid = {31086298},
publisher = {Springer US},
title = {{Long-range enhancer–promoter contacts in gene expression control}},
url = {http://dx.doi.org/10.1038/s41576-019-0128-0},
volume = {20},
year = {2019}
}

@article{Grah2020a,
author = {Grah, Rok and Zoller, Benjamin and Tka{\v{c}}ik, Ga{\v{s}}per},
doi = {10.1101/2020.04.08.029405},
journal = {bioRxiv},
pages = {2020.04.08.029405},
publisher = {Cold Spring Harbor Laboratory},
title = {{Normative models of enhancer function}},
url = {https://doi.org/10.1101/2020.04.08.029405},
year = {2020}
}

@article{Scholes2017,
author = {Scholes, Clarissa and Depace, Angela H and {Lvaro S{\'{a}} Nchez}, A ´},
doi = {10.1016/j.cels.2016.11.012},
journal = {Cell Systems},
pages = {97--108},
title = {{Math j Bio Combinatorial Gene Regulation through Kinetic Control of the Transcription Cycle}},
url = {http://dx.doi.org/10.1016/j.cels.2016.11.012http://dx.doi.org/10.1016/j.cels.2016.11.012},
volume = {4},
year = {2017}
}

@article{Li2018b,
author = {Li, Congxin and Cesbron, Fran{\c{c}}ois and Oehler, Michael and Brunner, Michael and H{\"{o}}fer, Thomas},
doi = {10.1016/j.cels.2018.01.012},
issn = {24054712},
journal = {Cell Systems},
pages = {409-423},
pmid = {29454937},
title = {{Frequency Modulation of Transcriptional Bursting Enables Sensitive and Rapid Gene Regulation}},
url = {http://linkinghub.elsevier.com/retrieve/pii/S2405471218300127},
year = {2018},
volume = {6}
}

@article{Furlong2018,
author = {Furlong, Eileen E.M. and Levine, Michael},
doi = {10.1126/science.aau0320},
issn = {10959203},
journal = {Science},
number = {6409},
pages = {1341--1345},
pmid = {30262496},
title = {{Developmental enhancers and chromosome topology}},
volume = {361},
year = {2018}
}

@article{Shao2017,
author = {Shao, Wanqing and Zeitlinger, Julia},
doi = {10.1038/ng.3867},
issn = {15461718},
journal = {Nature Genetics},
number = {7},
pages = {1045--1051},
pmid = {28504701},
publisher = {Nature Publishing Group},
title = {{Paused RNA Polymerase II inhibits new transcriptional initiation}},
volume = {49},
year = {2017}
}

@article{Gebhardt2013,
author = {Gebhardt, J. Christof M. and Suter, David M. and Roy, Rahul and Zhao, Ziqing W. and Chapman, Alec R. and Basu, Srinjan and Maniatis, Tom and Xie, X. Sunney},
doi = {10.1038/nmeth.2411},
page = {421--426},
volume = {10},
issn = {15487091},
journal = {Nature Methods},
title = {{Single-molecule imaging of transcription factor binding to DNA in live mammalian cells}},
year = {2013}
}

@article{Shelansky2020b,
author = {Shelansky, Robert and Boeger, Hinrich},
doi = {10.1073/PNAS.1911188117},
issn = {0027-8424},
journal = {Proceedings of the National Academy of Sciences of the United States of America},
number = {5},
pages = {2456--2461},
pmid = {31964832},
publisher = {National Academy of Sciences},
title = {{Nucleosomal proofreading of activator–promoter interactions}},
url = {https://www.pnas.org/content/117/5/2456},
volume = {117},
year = {2020}
}

@article{Kouzarides2007,
author = {Kouzarides, Tony},
doi = {10.1016/j.cell.2007.02.005},
journal = {Cell},
number = {4},
pages = {693--705},
pmid = {17320507},
title = {{Chromatin modifications and their function}},
volume = {128},
year = {2007}
}

@article{Benabdallah2019,
author = {Benabdallah, Nezha S. and Williamson, Iain and Illingworth, Robert S. and Kane, Lauren and Boyle, Shelagh and Sengupta, Dipta and Grimes, Graeme R. and Therizols, Pierre and Bickmore, Wendy A.},
doi = {10.1016/j.molcel.2019.07.038},
issn = {10974164},
journal = {Molecular Cell},
number = {3},
pages = {473--484.e7},
pmid = {31494034},
publisher = {Elsevier Inc.},
title = {{Decreased enhancer-promoter proximity accompanying enhancer activation}},
url = {https://doi.org/10.1016/j.molcel.2019.07.038},
volume = {76},
year = {2019}
}

@article{Estrada2016a,
author = {Estrada, Javier and Wong, Felix and DePace, Angela and Gunawardena, Jeremy},
doi = {10.1016/j.cell.2016.06.012},
issn = {1097-4172},
journal = {Cell},
number = {1},
pages = {234--44},
pmid = {27368104},
publisher = {NIH Public Access},
title = {{Information Integration and Energy Expenditure in Gene Regulation.}},
volume = {166},
year = {2016}
}

@article{Bothma2014,
author = {Bothma, Jacques P. and Garcia, Hernan G. and Esposito, Emilia and Schlissel, Gavin and Gregor, Thomas and Levine, Michael},
doi = {10.1073/pnas.1410022111},
issn = {10916490},
journal = {Proceedings of the National Academy of Sciences},
number = {29},
pages = {10598--10603},
publisher = {National Academy of Sciences},
title = {{Dynamic regulation of eve stripe 2 expression reveals transcriptional bursts in living Drosophila embryos}},
volume = {111},
year = {2014}
}

@article{Teves2018,
author = {Teves, Sheila S. and An, Luye and Bhargava-Shah, Aarohi and Xie, Liangqi and Darzacq, Xavier and Tjian, Robert},
doi = {10.7554/eLife.35621},
issn = {2050084X},
journal = {eLife},
pages = {e35621},
pmid = {29939130},
title = {{A stable mode of bookmarking by TBP recruits RNA polymerase II to mitotic chromosomes}},
volume = {7},
year = {2018}
}

@article{Krebs2017,
author = {Krebs, Arnaud R. and Imanci, Dilek and Hoerner, Leslie and Gaidatzis, Dimos and Burger, Lukas and Sch{\"{u}}beler, Dirk},
doi = {10.1016/j.molcel.2017.06.027},
issn = {10974164},
journal = {Molecular Cell},
number = {3},
pages = {411--422.e4},
publisher = {Cell Press},
title = {{Genome-wide single-molecule footprinting reveals high RNA Polymerase II turnover at paused promoters}},
volume = {67},
year = {2017}
}

@article{Cisse2013,
author = {Cisse, Ibrahim I. and Izeddin, Ignacio and Causse, Sebastien Z. and Boudarene, Lydia and Senecal, Adrien and Muresan, Leila and Dugast-Darzacq, Claire and Hajj, Bassam and Dahan, Maxime and Darzacq, Xavier},
doi = {10.1126/science.1239053},
issn = {10959203},
journal = {Science},
number = {6146},
pages = {664--667},
title = {{Real-time dynamics of RNA polymerase II clustering in live human cells}},
volume = {341},
year = {2013}
}

@article{Morrison2020,
author = {Morrison, Muir and Razo-Mejia, Manuel and Phillips, Rob},
doi = {10.1101/2020.06.13.150292},
journal = {bioRxiv},
pages = {2020.06.13.150292},
publisher = {Cold Spring Harbor Laboratory},
title = {{Reconciling Kinetic and Equilibrium Models of Bacterial Transcription}},
url = {https://doi.org/10.1101/2020.06.13.150292},
year = {2020}
}

@article{Ghuysen2007,
author = {Ghuysen, J M and Gange, D and Devriese, L A and Haesebrouck, F and Sander, C and Mckeveney, D and Muldoon, C and Rajaratnam, P and Meutermans, W and Duewel, H S and Honek, J F and Berghuis, A M and Chen, L and Litterman, N K and Walker, S and Fastrez, J and Declercq, J P},
journal = {Science},
number = {March},
pages = {1405--1409},
title = {{Dynamics of replication-independent Histone turnover in budding yeast}},
volume = {315},
year = {2007}
}

@article{Mir2017,
author = {Mir, Mustafa and Reimer, Armando and Haines, Jenna E. and Li, Xiao Yong and Stadler, Michael and Garcia, Hernan and Eisen, Michael B. and Darzacq, Xavier},
doi = {10.1101/gad.305078.117},
issn = {15495477},
journal = {Genes and Development},
number = {17},
pages = {1784--1794},
publisher = {Cold Spring Harbor Laboratory Press},
title = {{Dense bicoid hubs accentuate binding along the morphogen gradient}},
volume = {31},
year = {2017}
}

@article{Core2008,
author = {Core, Leighton J. and Waterfall, Joshua J. and Lis, John T.},
doi = {10.1126/science.1162228},
issn = {00368075},
volume = {322},
pages = {1845-1848},
journal = {Science},
title = {{Nascent RNA sequencing reveals widespread pausing and divergent initiation at human promoters}},
year = {2008}
}

@article{McKnight1979,
author = {McKnight, Steven L. and Miller, Oscar L.},
doi = {10.1016/0092-8674(79)90263-0},
issn = {00928674},
journal = {Cell},
number = {3},
pages = {551--563},
title = {{Post-replicative nonribosomal transcription units in D. melanogaster embryos}},
volume = {17},
year = {1979}
}

@article{Bartman2019a,
author = {Bartman, Caroline R. and Hamagami, Nicole and Keller, Cheryl A. and Giardine, Belinda and Hardison, Ross C. and Blobel, Gerd A. and Raj, Arjun},
doi = {10.1016/j.molcel.2018.11.004},
issn = {10974164},
journal = {Molecular Cell},
number = {3},
pages = {519--532.e4},
pmid = {30554946},
publisher = {Elsevier Inc.},
title = {{Transcriptional burst initiation and Polymerase pause release are key control points of transcriptional regulation}},
url = {https://doi.org/10.1016/j.molcel.2018.11.004},
volume = {73},
year = {2019}
}

@article{Chen2014,
author = {Chen, Jiji and Zhang, Zhengjian and Li, Li and Chen, Bi Chang and Revyakin, Andrey and Hajj, Bassam and Legant, Wesley and Dahan, Maxime and Lionnet, Timoth{\'{e}}e and Betzig, Eric and Tjian, Robert and Liu, Zhe},
doi = {10.1016/j.cell.2014.01.062},
issn = {10974172},
journal = {Cell},
number = {6},
pages = {1274--1285},
publisher = {Cell Press},
title = {{Single-molecule dynamics of enhanceosome assembly in embryonic stem cells}},
volume = {156},
year = {2014}
}

@article{Wong2020,
author = {Wong, Felix and Gunawardena, Jeremy},
doi = {10.1146/annurev-biophys-121219},
title = {Gene Regulation in and out of Equilibrium},
journal = {Annual Review of Biophysics},
volume = {49},
pages = {199--226},
year = {2020}
}

@article{Sanchez2013,
author = {Sanchez, Alvaro and Golding, Ido},
journal = {Science},
doi = {10.1126/science.1242975},
issn = {10959203},
number = {6163},
pages = {1188--1193},
pmid = {24311680},
publisher = {American Association for the Advancement of Science},
title = {{Genetic determinants and cellular constraints in noisy gene expression}},
url = {https://science.sciencemag.org/content/342/6163/1188 https://science.sciencemag.org/content/342/6163/1188.abstract},
volume = {342},
year = {2013}
}

@article{Rodriguez2020,
author = {Rodriguez, Joseph and Larson, Daniel R and ARjatscls, Larson},
doi = {10.1146/annurev-biochem-011520},
pages = {1--24},
title = {{Transcription in Living Cells: Molecular Mechanisms of Bursting}},
url = {https://doi.org/10.1146/annurev-biochem-011520-},
journal = {Annual Review of Biochemistry},
year = {2020}
}

@article{Desponds2016e,
author = {Desponds, Jonathan and Tran, Huy and Ferraro, Teresa and Lucas, Tanguy and {Perez Romero}, Carmina and Guillou, Aurelien and Fradin, Cecile and Coppey, Mathieu and Dostatni, Nathalie and Walczak, Aleksandra M.},
doi = {10.1371/journal.pcbi.1005256},
issn = {1553-7358},
journal = {PLOS Computational Biology},
number = {12},
title = {{Precision of readout at the hunchback gene: analyzing short transcription time traces in living fly embryos}},
url = {http://dx.plos.org/10.1371/journal.pcbi.1005256},
volume = {12},
year = {2016}
}

@article{Bowman2015,
author = {Bowman, Gregory D. and Poirier, Michael G.},
doi = {10.1021/cr500350x},
journal = {Chemical Reviews},
number = {6},
pages = {2274--2295},
pmid = {25424540},
title = {{Post-translational modifications of histones that influence nucleosome dynamics}},
volume = {115},
year = {2015}
}

@article{Mir2018b,
author = {Mir, Mustafa and Stadler, Michael R. and Ortiz, Stephan A. and Hannon, Colleen E. and Harrison, Melissa M. and Darzacq, Xavier and Eisen, Michael B.},
doi = {10.7554/eLife.40497},
issn = {2050084X},
journal = {eLife},
number = {Dv},
pages = {1--27},
pmid = {30589412},
title = {{Dynamic multifactor hubs interact transiently with sites of active transcription in drosophila embryos}},
volume = {7},
year = {2018}
}

@article{Braun2017,
author = {Braun, Simon M.G. and Kirkland, Jacob G. and Chory, Emma J. and Husmann, Dylan and Calarco, Joseph P. and Crabtree, Gerald R.},
doi = {10.1038/s41467-017-00644-y},
isbn = {4146701700644},
issn = {20411723},
journal = {Nature Communications},
number = {1},
publisher = {Springer US},
title = {{Rapid and reversible epigenome editing by endogenous chromatin regulators}},
url = {http://dx.doi.org/10.1038/s41467-017-00644-y},
volume = {8},
year = {2017}
}

@article{Xu2015b,
author = {Xu, Heng and Sep{\'{u}}lveda, Leonardo A. and Figard, Lauren and Sokac, Anna Marie and Golding, Ido},
doi = {10.1038/nmeth.3446},
issn = {15487105},
journal = {Nature Methods},
number = {8},
pages = {739--742},
publisher = {Nature Publishing Group},
title = {{Combining protein and mRNA quantification to decipher transcriptional regulation}},
volume = {12},
year = {2015}
}

@article{Park2019,
author = {Park, Jeehae and Estrada, Javier and Johnson, Gemma and Vincent, Ben J and Ricci-tam, Chiara and Bragdon, Meghan D J and Shulgina, Yekaterina and Cha, Anna and Wunderlich, Zeba and Gunawardena, Jeremy and Depace, Angela H},
pages = {1--25},
title = {{Dissecting the sharp response of a canonical developmental enhancer reveals multiple sources of cooperativity}},
year = {2019},
journal = {eLife},
volume = {8},
pages = {e41266}
}

@article{Halpern2015,
author = {Halpern, Keren Bahar},
doi = {10.1016/j.molcel.2015.01.027},
journal = {Molecular Cell},
pages = {147--156},
title = {{Bursty gene expression in the intact mammalian liver}},
url = {http://dx.doi.org/10.1016/j.molcel.2015.01.027},
volume = {58},
year = {2015}
}

\end{refsection}

\end{document}